\documentclass[american,aps,pra,reprint, superscriptaddress]{revtex4-2}
\usepackage{color}
\usepackage{times}
\usepackage{subfig}
\usepackage{bm}
\usepackage{graphicx}
\usepackage{graphics}
\DeclareGraphicsExtensions{.pdf,.png,.jpg}
\usepackage{amsbsy}
\usepackage{amsmath}
\usepackage{amsfonts}
\usepackage{amsthm}
\usepackage{float}
\usepackage{amssymb}
\usepackage{textcomp}
\usepackage[subnum]{cases}
\graphicspath{{fig/}}

\definecolor{purple(html/css)}{rgb}{0.5, 0.0, 0.5}
\newcommand{\ket}[1]{| #1 \rangle}
\newcommand{\bra}[1]{\langle #1 |}

\newcommand{\ketbra}[2]{| #1 \rangle \langle #2 |}

\newcommand{\mi}{\mathrm{i}}
\begin{document}
\title{Creation of quantum coherence with general measurement processes}

\author{Sanuja D. Mohanty}
%\email{sanujamohanty93@gmail.com}
\affiliation{International Institute of Information Technology, Bhubaneswar, 751003, India}

\author{Gautam Sharma}
\email{gautam.oct@gmail.com}
\affiliation{Optics and Quantum Information Group, Institute of Mathematical Sciences, HBNI, Chennai 600113, India}

\author{Sk Sazim}
\email{sk.sazimsq49@gmail.com}
\affiliation{Harish-Chandra Research Institute, HBNI, Allahabad, 211019, India}
\affiliation{RCQI, Institute of Physics, Slovak Academy of Sciences, 845 11 Bratislava, Slovakia}

\author{Biswajit Pradhan}
\affiliation{International Institute of Information Technology, Bhubaneswar, 751003, India}

\author{Arun K. Pati}
\email{akpati@hri.res.in}
\affiliation{Harish-Chandra Research Institute, HBNI, Allahabad, 211019, India}

\date{\today}

\begin{abstract}
Quantum measurement not only can destroy coherence but also can create it. Here, we estimate 
%ask how much 
the maximum amount of coherence, one can create under a complete non-selective measurement process. For our analysis, we consider  
%We find answers to this question for 
projective as well as POVM measurements. Based on our observations, we characterize the measurement processes into two categories, namely, the measurements with the ability to induce coherence and the ones without this ability. Our findings also suggest that the more POVM elements present in a measurement that acts on the quantum system, the less will be its coherence creating ability. We also introduce the notion of raw coherence in the POVMs that helps to create quantum coherence. Finally, we find a trade-off relation between the coherence creation, entanglement generation between system and apparatus, and the mixedness of the system in a general measurement setup.

\end{abstract}

\maketitle
		
%\tableofcontents
	
\section{Introduction}
The ‘superposition principle’ demarcates the quantum world from its classical counterpart. Quantum coherence is a resource that arises due to the superposition principle \cite{J_Aberg_Coh}. Recently, a rigorous analysis of quantum coherence as a resource has been done in \cite{Baumg_Cohere}. This promulgates the wellknown non-classical resource, the coherence, in the forefront of quantum information science \cite{Winter_Cohere, Streltsov_Coh,Wu2020,PhysRevA.102.050202,2020arXiv201107239K,PhysRevA.103.032429}. As quantum coherence depends on the basis in consideration, it can be created and destroyed by unitary operations. For example, a simple rotation on a Bloch sphere may create a unit amount of coherence from an incoherent state. This gives the hint that there exists a class of quantum operations that create or destroy quantum coherence – this power of quantum operations is respectively known as cohering and decohering power \cite{Mani_CP,2015arXiv151006683G,PhysRevA.95.052306,PhysRevA.95.052307,Bu_CP,PhysRevA.97.032304,2018QuIP...17..186Z,PhysRevA.94.012326}. 
%%%%%%%%%%%%%%%%%%%%%%%%%%%%%%%%%%%%%%%%%%%%%%%%%%%%%%%%%%%%%%%

%We have general consensus that quantum operations usually destroy quantum resources in quantum systems \cite{NielsonBook}. 
It is understandable that quantum coherence in the quantum states gets destroyed when we perform projective measurement on an incoherent basis \cite{NielsonBook}. A selective projective measurement on a coherent basis can always create coherence. However, this might not be the case when we are not selecting any outcome. This becomes more non-trivial when we are consider positive operator value measurement (POVMs). For example, consider the action of a POVM  with elements $\{ M_i = \ketbra{+}{\psi_i}, M_0 = \mathbb{I}- \sum_i M_i\}$. For this POVM, $\forall i$, the instrument ignores the input state and outputs a maximally coherent state, however, if we do not select any outcome, we get a different state. Therefore, we address the question: how much coherence one can create using general non-selective measurements. 

In this work, we investigate the cohering capability of general quantum operations allowed on qubits. This provides us a hint that the coherence, unlike other intricate quantum resources, eg., entanglement, may not be as vulnerable as was presumed earlier. Before going into the main discussion of the results, we list the key findings: (i) Non-selective measurement can induce coherence in the incoherent states. For arbitrary incoherent qubits states, we show that a POVM operation can induce coherence up to $|r_3|/2$, which can reach a maximum of 1/2, (ii) For every qubit state, it is not possible to induce coherence with a POVM, however we can still prevent the loss of coherence, (iii) The more elements in the POVM  (i.e., the measurement is becoming fuzzier) the less will be its coherence creation ability. The numerical result suggests that the induced coherence by $n$-outcome POVM is proportional to $e^{-bn}$, where $b\approx 0.37$, (iv) We coin a term called ‘raw quantumness ($C_{raw}$)’ for the elements of measurement. It is the sum of the norm of off-diagonal elements present in the POVM elements. We find that if all the POVM elements have $C_{raw} = 0$, its coherence creation ability is zero, (v) We characterize measurements in two categories – the coherence non-generating measurements and the coherence generating one and find their properties. Further, we also prove a trade-off relation between the coherence creation, entanglement between system and apparatus, and the mixedness of the system
	
The paper is organized as follows. In the prelude, a brief description about the resource theory of coherence are presented. In Sec.\ref{Effect of General Measurement on Coherence}, we address the question: how to create quantum coherence under complete measurement. 
% The answer to the question: \emph{can measurement creates coherence ?} is discussed in Sec.\ref{Effect of General Measurement on Coherence}. 
While, more specifically the role of general measurement processes on quantum coherence is presented in Sec.\ref{Role of POVM Operator on Quantum Coherence}. Some numerical and theoretical discussion and the validation of the proposed Hypothesis is discussed with figures. In Sec.\ref{new-coh-ent}, we link two well-known resources, namely, coherence and entanglement via generalized measurement scheme. Finally, we conclude in the Sec.\ref{Conclusion} with some future avenue of research.
	
\emph{Prelude}.-
%\label{Resource theory of coherence}
Any legitimate resource theory has two basic elements -- free states and free operations. For the resource theory of coherence, they are incoherent states and incoherent operations respectively. As this particular resource theory is basis dependent, we need to fix a basis. Let us consider the computational basis, $\{\ket{i}; i\in\mathbb{Z}^+\}$ in Hilbert space $\mathcal{H}$, with $|\mathbb{Z}^+|={\rm dim}(\mathcal{H})$, where $\mathbb{Z}^+$ is set of non-negative integers. 
The diagonal density matrices in this basis are incoherent states and expressed as 
\begin{equation}
\label{eqa}
\delta=\sum_{i\in\mathbb{Z}^+} \delta_{ii}\ket{i}\bra{i}.
\end{equation}
The set of incoherent states is represented by $\mathcal{I}$. The operations which keeps all incoherent states incoherent, are called incoherent operations. 
	
The quantification of resource is an important aspect for its physical implications. Before going into the measure of coherence, we recall from literature that what are the basic requirements for such functions to be valid measure of coherence. The following properties a function should satisfy to be a valid measure of quantum coherence \cite{Streltsov_Coh}:\\
% \begin{enumerate}
C1). Coherence vanishes for all incoherent state, $C(\delta)=0$ for all  $\delta\in\mathcal{I}$.\\
C2). Coherence should not increase under mixing of states, i.e., $\sum_ip_iC(\rho_i)\geq C(\sum_ip_i\rho_i)$.\\
C3a). Monotonicity under incoherent completely positive and trace preserving (CPTP) maps, $\Phi$: $C(\rho)\geq C(\Phi[\rho])$.\\
C3b). Monotonicity under selective incoherent operations on average $C(\rho)\geq\sum_{i} p_nC(\rho_n)$, where $p_n= {\rm Tr}[{K_n}\rho K_n^{\dagger}]$, and $\rho_n=\frac{1}{p_n}{K_n}\rho K_n^{\dagger}$ with $\{ K_n \} $ is Kraus decomposition of $\Phi$.
% \end{enumerate}
	
There exist several quantum coherence measures \cite{Streltsov_Coh}. However, we will focus on the $l_1$-norm of coherence and the relative entropy of coherence 
\cite{Baumg_Cohere}. The $l_1$ norm of coherence is defined as 
\begin{equation}
\label{eqc}
\ C_{l_1}(\rho)=\sum_{i\neq j}|\rho_{i,j}|,
\end{equation}
where $|X|$ denotes absolute value of $X$. This measure captures the off-diagonal elements of a density matrix, and thus has very important physical implications. For example, using the $l_1$-norm coherence, a
duality relation between coherence and the path information has been proved \cite{math4030047,PhysRevA.92.042101}.
	
The relative entropy of coherence is defined as 
\begin{equation}
\label{eqd}
\ C_{r}(\rho)=S(\rho||\rho^D)=S(\rho^D)-S(\rho),
\end{equation}
where $S(\rho)=-{\rm Tr}[\rho\log_2\rho]$ is von Neumann entropy and $\rho^D=\sum_i\bra{i}\rho\ket{i}\ketbra{i}{i}$ is completely dephasing of $\rho$. Geometrically, it 
is saying that how far an arbitrary state is from its closest incoherent state. This has a beautiful physical implication as it quantifies exactly the amount of distillable 
coherence from a mixed quantum state \cite{Winter_Cohere}. Also, it has a nice thermodynamic meaning \cite{PhysRevLett.119.150405}.

In the initial calculations involving study of increment in coherence under specific qubit operations we will use the $l_1$-norm of coherence as the calculations are easily doable. Later, we will use the relative entropy of coherence as it appears naturally in certain specific cases.
	
\section{Creation of quantum coherence by a measurement}
\label{Effect of General Measurement on Coherence}
Any measurement can mathematically be represented by a set of Positive-Operator-Valued-Measures (POVM), i.e., $\{E_i; \forall i\in \mathbb{Z}^+\}$, with $E_i\geq 0$ and 
$\sum_iE_i=\mathbb{I}$ \cite{VonNeumann, measurementWheeler}. The action of this measurement on a quantum state $\rho$ changes it to $\rho^M=\sum_i\sqrt{E_i}\rho\sqrt{E_i}$ 
in non-selective case, and for selective case, it results a post measurement state $\rho_i^M=\frac{1}{p_i}\sqrt{E_i}\rho\sqrt{E_i}$ 
with probability $p_i={\rm Tr}[E_i\rho]$ \cite{1983LNP...190.....K, busch-etal:1995,2006AnP...518..663L}. 
Any projective measurement is a subset of the above general measurements. A projective measurement consists of set orthogonal effects, 
$\{\Pi_i; i=0,1,..., d\}$, where $d$ is the dimension of the underlying Hilbert space \cite{VonNeumann}.

Let us consider an arbitrary single qubit state, $\ket{\psi}=\alpha \ket{0}+  \sqrt{1-|\alpha|^2}\ket{1}$, where $\alpha \in \mathcal{C}$.  
Under non-selective projective measurement with projectors, $\{\Pi_0=\ketbra{0}{0},\Pi_1=\ketbra{1}{1}\}$, we have  
\begin{eqnarray}\label{eqe}
    \rho^M &=& \sum_{i=0}^1\Pi_i ~ \ket{\psi} \bra{\psi} ~ \Pi_i,\nonumber\\
     &=&|\alpha|^2\ket{0} \bra{0} + (1-|\alpha|^2)\ket{1} \bra{1}.
\end{eqnarray}
The $l_1$-norm of coherence of the states before and after measurement in $\{\ket{0},\ket{1}\}$ basis can be presented as 
\begin{eqnarray}
    C_{l_1}(\ket{\psi}) &=& 2|\alpha||\sqrt{1-\alpha^2}|,\nonumber\\
\mbox{and} ~ && C_{l_1}(\rho^M)=0.
\end{eqnarray}
Therefore, it is clear that due to complete measurement, the initial state undergoes a dephasing and hence loses its all coherence. This very phenomenon is known as `de-coherence'. However, we will show that a complete projective measurement sometime can create coherence instead of destroying it.

% Next, we find that even though the unitary and the measurement processes are fundamentally different, still they can create same amount of coherence starting from an initially incoherent state, i.e, $\ket{0}$. The scenario is depicted below. 

Let us consider the incoherent state $\delta_0=\ket{0}\bra{0}$ and apply the measurement operator defined using the basis $\{\ket{\psi},\tilde{\ket{\psi}} \}$, where $\tilde{\ket{\psi}}= \sqrt{1-|\alpha|^2}\ket{0}- \alpha^*\ket{1}$ and $*$ denotes complex conjugation. The final state can be presented as
\begin{eqnarray}
\label{eqg}
\rho^M\to \bra{\psi} \delta_0 \ket{\psi} \ket{\psi} \bra{\psi} + \tilde{\bra{\psi}} \delta_0 \tilde{\ket{\psi}} \tilde{\ket{\psi}} \tilde{\bra{\psi}}\nonumber\\
\rho^M=|\alpha|^2\ket{\psi}\bra{\psi}+(1-|\alpha|^2)\tilde{\ket{\psi}}\tilde{\bra{\psi}}
\end{eqnarray}	
The ${l_1}$ norm coherence of the final state in $\{\ket{0},\ket{1}\}$ basis is given by 
\begin{eqnarray}
\label{eqh}
C_{l_1}(\rho^M)=2|\alpha|\sqrt{1-|\alpha|^2}(|2|\alpha|^2-1)|).
%=2||\alpha|^2-|\beta|^2| |\alpha||\beta|
\end{eqnarray}
We conclude that, non-selective measurement in $\{ \ket{\psi}, \tilde{\ket{\psi}} \}$ basis has created coherence for an incoherent state. The maximum value of $C_{l_1}(\rho^M)$ reaches to $\frac{1}{2}$ for $|\alpha|=(\sqrt{2\mp\sqrt{2}})/2$, i.e., $|\alpha|\approx 0.384$ and $0.924$.

% Any qubit incoherent state in $\{\ket{0},\ket{1}\}$ can be expressed as $\delta=x\ketbra{0}{0}+(1-x)\ketbra{1}{1}$ with $x\in [0,1]$. Then, the average coherence generated by the measurement $\{\Pi_0,\Pi_1\}$ from any incoherent state can be evaluated as 
% % 	
% % Now, we can choose an unitary operation, $U$, for transformation as 
% \begin{eqnarray}
% \label{eqi}
% C_{l_1}^{avg}&=&2|\alpha|\sqrt{1-|\alpha|^2}(|2|\alpha|^2-1)|)\int_{0}^1|2x-1|dx,\nonumber\\
% &=&|\alpha|\sqrt{1-|\alpha|^2}(|2|\alpha|^2-1)|).
% \end{eqnarray}
% Therefor, the maximum average coherence created by the above projective measurement from any incoherent state is $0.25$, which occurs when $|\alpha|=(\sqrt{2\mp\sqrt{2}})/2$.
% where $\ket{\psi_t}=\frac{1}{\sqrt{N}}(\alpha\sqrt{(1-|\alpha|^2)}\ket{0}+(2|\alpha|^2-1)\ket{1})$, where $N=3|\alpha|^2(|\alpha|^2-1)+1$. The coherence of the rotated state, $\ket{\psi_t}$ is 
% \begin{equation}
% \label{eqj}
%  C_{l_1}(\ket{\psi_t})=\frac{1}{N}C_{l_1}(\rho^M).
% \end{equation}
% This example is sufficient to conclude that; though, the unitary and measurement processes are different but they can create almost (scaled) same amount of coherence from an initial incoherent state.
% 	
\section{Role of generalized measurement on Quantum Coherence}
\label{Role of POVM Operator on Quantum Coherence}
Any generalized measurement process can be described as the action of $n$ element positive-operator valued measurements (POVM), i.e., $\{E_i;i=1,2,...,n\}$, with $\sum_{i=1}^n E_i=\mathbb{I}$. However, to describe the effect of quantum measurement on the quantum state, initially we will consider the scenarios where a limited number of POVM elements are present and only later we give numerical results for n element POVM.
	
\subsection{One-parameter POVM Operators}
\begin{figure}[htb]
\centering
\includegraphics[scale=0.65]{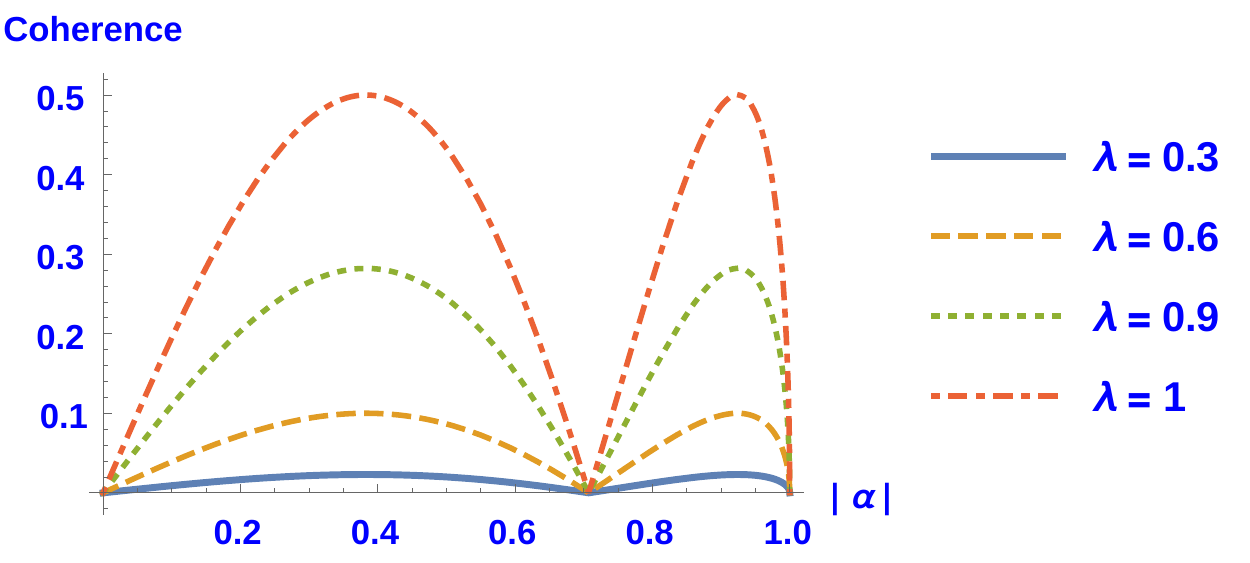}
\caption{(Color online) The coherence gain after the action of one-parameter POVM  measurement over incoherent state. It is showing that the more sharp the measurement is the more is the coherence gain. The curve, $\lambda=1$ depicts the Eq.(\ref{eqh}).}
% Note that there is no coherence gain at the point $|\alpha|=\frac{1}{\sqrt{2}}$.}
\label{fig:x}
\end{figure}

Let us consider the following one parameter POVM decomposition of a unsharp measurement
\begin{equation}
\label{eqm}
 E_{\pm}=\lambda P_{\pm}+\frac{1-\lambda}{2}\mathbb{I},
\end{equation}
where $\lambda$ is the sharpness parameter, $0\leq \lambda\leq 1$, $P_{+}=\ket{\psi}\bra{\psi}$, $P_{-}=\tilde{\ket{\psi}}\tilde{\bra{\psi}}$ and $E_++E_-=\mathbb{I}$. Now the final state, due to the measurement on $\delta_0=\ket{0}\bra{0}$, can be evaluated using L\"{u}der's rule \cite{2006AnP...518..663L},
\begin{eqnarray}
\label{eqo}
\delta_0'&=&
\sqrt{1-\lambda^2}\delta_0+(1-\sqrt{1-\lambda^2})(P_+\delta_0 P_++P_-\delta_0 P_-) \nonumber\\
&=&\sqrt{1-\lambda^2}\delta_0+(1-\sqrt{1-\lambda^2}) \rho^M,
\end{eqnarray}
where $\rho^M$ is defined in Eq. (\ref{eqg}). The $l_1$-norm coherence of the state $\delta_0'$  is
\begin{equation}
\label{eqp}
C_{{l_1}}(\delta_0')=(1-\sqrt{1-\lambda^2}) C_{l_1}(\rho^M).  
\end{equation}
This example shows that one can create a non-zero amount of coherence from an incoherent state due to POVM measurement. It is also clear that the more sharp is the measurement, the more will be the coherence creation (see Fig.\ref{fig:x}).

Now, if we consider the initial state to be the more general state with Bloch vector $\vec{r}$, i.e., $\rho=\frac{1}{2}(\mathbb{I} + \vec{r}.\vec{\sigma})$, the evolved state under above POVM measurement will transform to
\begin{equation}
\label{eqoG}
\rho'
%\sqrt{1-\lambda^2}\rho_{\pm}+(1-\sqrt{1-\lambda^2})(P_+\rho_{\pm} P_++P_-\rho_{\pm} P_-) \nonumber\\
=\sqrt{1-\lambda^2}\rho+(1-\sqrt{1-\lambda^2})(P_{+}\rho P_{+}+P_-\rho P_-).
\end{equation}
So, the coherence of the final state is given by
\begin{align}
 C_{l_1}(\rho')
 &=2\Big|(1-\sqrt{1-\lambda^2})\bar{\alpha}\Big[\alpha \bar{\alpha}(R^* \alpha-r_3\bar{\alpha})\nonumber \\  & +|\alpha|^2(r_3\alpha+R\bar{\alpha})\Big]+\frac{\sqrt{1-\lambda^2}}{2}R\Big|, 
 %\nonumber \\
% &+\Big|\frac{\sqrt{1-\lambda^2}}{2}R+(1-\sqrt{1-\lambda^2})\beta%%
%\Big(\alpha^* \beta((r_1+\mi r_2) \alpha^*-r_3\beta)\nonumber \\  & +|
%\alpha|^2(r_3\alpha^*+(r_1-\mi r_2)\beta)\Big)\Big|,
\end{align}
where $R=r_1+\mi r_2$ and $\bar{\alpha}=\sqrt{1-|\alpha|^2}$, and the coherence of the initial state is given by  $C_{l_1}(\rho)=|R|$. We are interested in whether there exist POVMs for which $C_{l_1}(\rho')>C_{l_1}(\rho)$. Therefore, we maximize $C_{l_1}(\rho')$ over the POVM parameters $\alpha$ and $\lambda$, which gives the following 
\begin{align}
{\max_{\alpha,\lambda}}C_{l_1}(\rho')&=\frac{1}{2}\big(\sqrt{r_1^2+r_2^2}+\sqrt{r_1^2+r_2^2+r_3^2}\big)\nonumber \\
&=\frac{1}{2}\big(C_{l_1}(\rho)+\sqrt{C_{l_1}(\rho)^2+r_3^2}\big).
\end{align}
Therefore, it can be seen that, we can have $\max_{\alpha,\lambda}C_{l_1}(\rho')> C_{l_1}(\rho)$, by choosing a suitable POVM. Also, it should be noted that this value is obtained for $\lambda=1$. i.e., maximally sharp measurements can induce more coherence in arbitrary qubit states. Where as at $\lambda =0$, in the case of trivial measurement of $\mathbb{I}/2$, we have $\max_{\alpha,\lambda}C_{l_1}(\rho')= C_{l_1}(\rho)$, as expected. For qubit states, in which $r_3=0$ there can't be any increment in the coherence of the state, however one can still ensure that there is no loss of coherence. 

Next important point to note is that for a given initial coherence $|R|$, ${\max_{\alpha,\lambda}}C_{l_1}(\rho')$ will have large value, only when $|r_3|$ is large. This happens for qubit states which have large $|\vec{r}|$, i.e., the states that are close to the surface of Bloch sphere, with the maximum value given by  ${\max_{\alpha,\lambda}}C_{l_1}(\rho')=\frac{1}{2}\big(C_{l_1}(\rho)+1\big)$ for pure qubit states.  It can thus be concluded that for two qubits with same initial coherence, the coherence gain will be more in the state which is more pure or has less mixedness.

\begin{figure}[htb]
\centering
\includegraphics[scale=0.65]{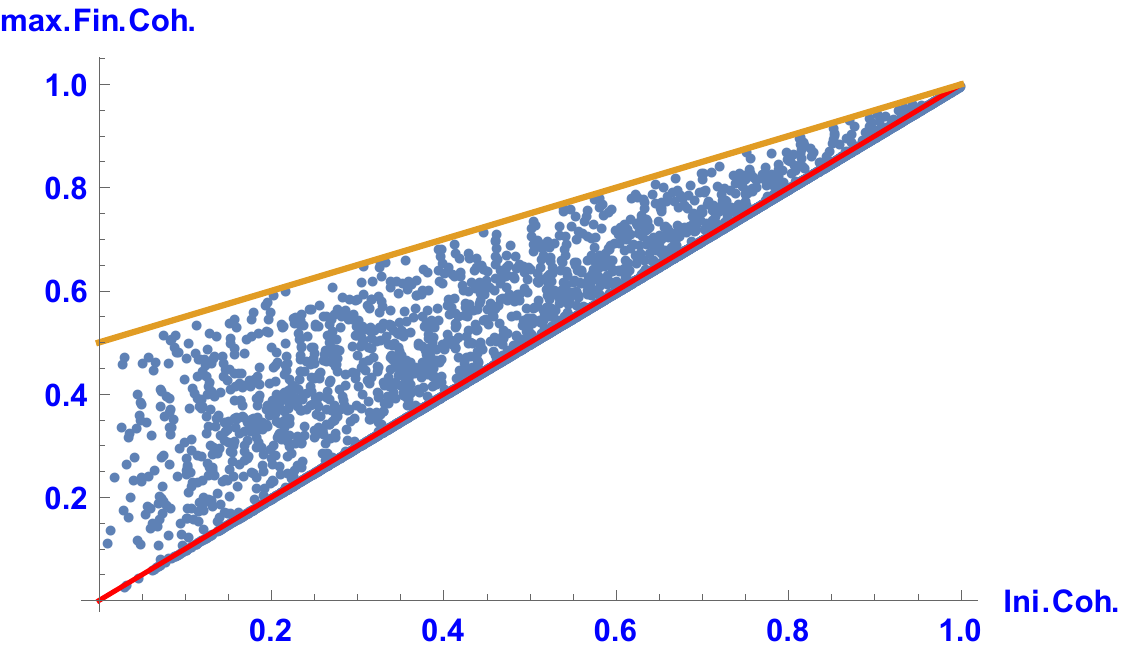}
\caption{(Color online) The plot of `$\max_{\{\lambda,\alpha\}} C_{l_1}(\rho'_{\pm})$ vs $C_{l_1}(\rho_{\pm})$' for one-parameter POVM. It shows many cases where the maximum final coherence is greater than the initial coherence. The red line depicts $\max_{\{\lambda,\alpha\}} C_{l_1}(\rho'_{\pm})$ = $C_{l_1}(\rho_{\pm})$ and the orange line which gives the upper bound depicts $\max_{\{\lambda,\alpha\}} C_{l_1}(\rho'_{\pm})$ = $\frac{C_{l_1}(\rho_{\pm})+1}{2}$. (The sample size is $15000$.)}
% Note that there is no coherence gain at the point $|\alpha|=\frac{1}{\sqrt{2}}$.}
\label{fig:y}
\end{figure}

Moreover, for an incoherent state $\rho$, i.e.,  with $|R|=0$, it can be seen that $\max_{\{\lambda,\alpha\}} C_{l_1}(\rho')=|r_3|/2$.  Hence, the maximum quantum coherence that we can create in an incoherent state is 1/2 which happens only for a pure incoherent state, i.e., with $|r_3|=1$. These observations can be also seen in Fig. \ref{fig:y}, where we numerically plot the $\max_{\{\lambda,\alpha\}} C_{l_1}(\rho')$ vs $C_{l_1}(\rho)$. The lower red line $y=x$ consists of states for which there is no increment in coherence whereas the orange line $y=\frac{x+1}{2}$ denotes the pure states for which maximum coherence gain happens. 
	%%%%%%%%%%%%%%%%%%%%%%%%%%%%%%%%%%%%%%%%%%%%%%%%%%%%%%%%%%%%%%%%%%%%%%%%%%%%%%%%%%%%%5
\subsection{General two outcome POVM Operators}
\label{Creating Coherence Using POVM Operator}

A more general form of POVM can be considered below
\begin{equation}\label{wer}
    E_{\pm}=a_{\pm} \mathbb{I} \pm \vec{a}.\vec{\sigma},
\end{equation}
where $\vec{a}$ is Bloch vector and $a_+ + a_- =1$, $0\leq a_{\pm}\leq 1$, $|a| \leq 
\min[a_+ ,a_- ]\leq \frac{1}{2}$. 
% One may implement this measurement (Eq.\ref{wer}) by $M_{\pm}=U\sqrt{E_{\pm}}$ where, $U$ is arbitrary unitary. 

Let us consider the evolution of the incoherent state $\delta_0$ under the above POVM measurement. The evolved state will be $\delta_0'=\sqrt{E_+}\delta_0\sqrt{E_+} +\sqrt{E_-}\delta_0\sqrt{E_-}$, i.e., $\delta_0' =\frac{1}{2}(\mathbb{I} \pm \vec{s}.\vec{\sigma})$, where, $\vec{s}$ is the evolved Bloch vector with elements, 
\begin{equation}
s_i=\gamma_0 a_i\:\: (i=1,2)\:\: \mbox{and} \:\:s_3=\frac{1}{|a|^2}\{a_3^2+(1-a_3^2)\beta\},
\end{equation}
where $\gamma_0=\frac{a_3}{|a|^2}[1-\beta]$, $\beta=\eta_++\eta_-$, and  $\eta_{\pm}=\sqrt{a_{\pm}^2-|a|^2}$. The coherence of the evolved state is 
\begin{equation}
    C_{l_1}(\delta_0')=\frac{|a_3|}{|a|^2}(|1-\beta|)\sqrt{a_1^2+a_2^2}.
    \label{sdf}
\end{equation}
Again, this example tells us that starting from zero coherence one can obtain a non-zero coherent state. It can be shown that  
\begin{align*}
	C_{l_1}(\delta_0')\leq\frac{|a_3|\sqrt{a_1^2+a_2^2}}{|a|^2}\Big(1-\sqrt{1-\sqrt{1-2|a|}}\Big)\leq 0.5.
\end{align*}
On maximizing over  $a_{\pm}$ and $a_i$'s, the maximum coherence one can create using this strategy from $\delta_0$ is $1/2$.

To complete our analysis, we consider the evolution of an arbitrary density matrix under generalized two-outcome POVM. An arbitrary density matrix, $\rho_{\pm}$ will evolve to 
	%$\rho'=\sqrt{E_+}\rho\sqrt{E_+} +\sqrt{E_-}\rho\sqrt{E_-}$, i.e., 
$\rho'_{\pm} =\frac{1}{2}(\mathbb{I} \pm \vec{s}.\vec{\sigma})$,
where, $\vec{s}$ is the evolved Bloch vector with elements, 
\begin{equation}
s_{\pm}^i=\theta_{\pm} a_i \pm\beta r_i, 
\end{equation}
with $\theta_{\pm}=\frac{\vec{a}.\vec{r}}{|a|^2}[\pm 1\mp\beta]$.
%, $\beta=\eta_++\eta_-$, and  $\eta_{\pm}=\sqrt{a_{\pm}^2-|a|^2}$. 
Then, the Bloch vector has been translated along with a rotation, i.e., $\vec{s}_{\pm}=\theta_{\pm} \vec{a} \pm\beta \vec{r}$. 
Now, the $l_1$-norm coherence of the state $\rho'_{\pm}$  is
\begin{equation}
\label{eqw}
C_{l_1}(\rho'_{\pm})=\left|\theta_{\pm} (a_1+\mi a_2) \pm\beta R\right|.
\end{equation}
	%where $\mu=\sqrt{\gamma \left(\frac{|a|^2-a_3^2}{|r|^2-r_3^2}\right)+2\gamma\beta \left(\frac{\vec{a}.\vec{r}-a_3r_3}{|r|^2-r_3^2}\right)+\beta}$. 
Note that the coherence of the initial state is $C_{l_1}(\rho_{\pm})=|R|$. 
%\begin{figure}[htb]
%\centering
%\includegraphics[scale=0.65]{povm_coh3.pdf}
%\caption{(Color online) The plot of `$\max_{\{\vec{a},a_{\pm}\}} C_{l_1}(\rho'_{\pm})$ vs $C_{l_1}(\rho_{\pm})$' for generalized two-outcome POVM. It shows many scenarios where the maximum final coherence is greater than the initial coherence. The red dashed line depicts $\max_{\{\lambda,\alpha\}} C_{l_1}(\rho'_{\pm})$ = $C_{l_1}(\rho_{\pm})$. (The sample size is $1000$.)}
%% Note that there is no coherence gain at the point $|\alpha|=\frac{1}{\sqrt{2}}$.}
%\label{fig:z}
%\end{figure}
Again, we are interested to see if there exists two outcome POVMs for which $C_{l_1}(\rho'_{\pm})>C_{l_1}(\rho)$. As before we can maximize over all the POVM parameters. For all the incoherent states, i.e., quantum states with $|R|=0$ we get, $\max_{\{\vec{a},a_{\pm}\}}C_{l_1}(\rho'_{\pm})=|r_3|/2\in [0,1/2]$. It can also be seen that, we can always have $C_{l_1}(\rho'_{\pm})=C_{l_1}(\rho)$ for $a_+=1/2$ and $a_i=0$ which is the trivial measurement with $\mathbb{I}/2$.
We couldn't get the expression for $\max_{\{\vec{a},a_{\pm}\}}C_{l_1}(\rho'_{\pm})$ analytically, so we have plotted it against the initial state coherence $C_{l_1}(\rho_{\pm})$. In this case also, we get same plot as in Fig.(\ref{fig:y}), i.e.,  a set of points bounded between lines, $y=x$ and $y=\frac{x+1}{2}$. Therefore, the one parameter POVM and the general two outcome POVM have similar ability to generate coherence. However in this case, we don't know yet for which states there can not be any coherence gain and for which states the coherence gain will be maximum.

%in Fig. \ref{fig:z}. The figure shows that there are many events where there is increase of coherence in the final state due to the generalized two-outcome POVM measurement.

\subsection{Randomly generated POVM operators with $n$ outcomes}
To generalize our study, we consider here the effect of $n$-outcome POVM on incoherent state $\delta_0$ in a two dimensional Hilbert space. Any $n$-outcome qubit POVM can be written as
\begin{align}\label{noutcomepovm}
E_i=a_i(\mathbb{I}+\vec{s}_i.\vec{\sigma}),	
\end{align}
with $a_i\geq 0$, $\sum_{i=1}^na_i=1$ and $\sum_ia_i\vec{s}_i=0$. However, we abstain ourselves from analytical results because of large number of parameters. We will use numerical simulations here to depict our findings. One can numerically generate $n$-outcome POVMs using QETLAB \cite{qetlab} or other method \cite{2019arXiv190204751H}. However, we will use QETLAB for our analysis.

\begin{figure}[htb]
	\centering
	\includegraphics[scale=0.45]{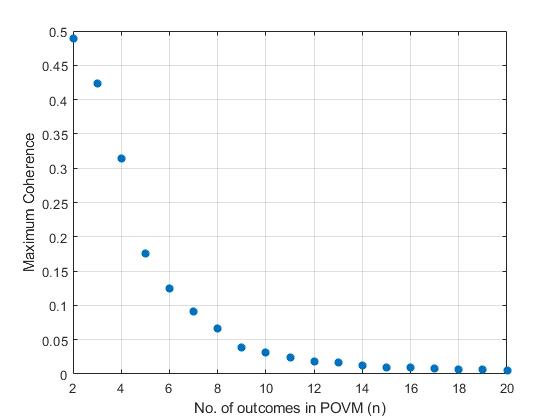}
	\caption{Maximum coherence gain after application of $n$-outcome randomly generated POVM measurement on incoherent state $\delta_0$. To achieve the numerical maximum, we have created $2.2\times 10^5$ of random POVM using QETLAB for each $n$. The plot shows that the maximum gain of coherence is decaying exponentially as we increase $n$.}
	\label{fig:e}
\end{figure}

We generate a sufficient amount of POVM sets ($2.2\times 10^5$) for each $n$ values to find the (almost)-maximum coherence state generated in the simulations. Then, we plot the maximum achievable coherence ($C_{\max}^n$) vs $n$ in the Fig.\ref{fig:e}. We restrict ourselves for $n\leq 20$. The Fig.\ref{fig:e}, shows that the maximum coherence generated from the incoherent state, using $n$-outcome random POVM is decreasing exponentially with $n$, i.e., $C_{\max}^n\propto e^{-bn}$ where $b\approx 0.37$. This behavior shows that the more the number of elements in the POVM sets, the less will be its coherence creation ability. This means the measurement is becoming more fuzzy. However, it is still unknown why this exponential behavior occurs.The above numerical observation suggests that the contribution to the induced coherence from the individual POVM elements decreases as $n$ increases, which motivates us to give the following conjecture.

{\em Conjecture}.-- The quantum coherence induced by an $n$-outcome POVM measurement decreases with increasing $n$.

%To support our claim, we will consider the following $n$-outcome POVM measurement 
%\begin{equation}
%	E_i=\frac{n_i}{n}(\mathbb{I}+\vec{s}_i.\vec{\sigma}),
%\end{equation}
%where $\sum_in_i=n$ and $\sum_in_i\vec{s}_i=0$. Now if we consider the action of this measurement on the incoherent state $\delta_0$, we will end up with the unnormalized $i^{th}$ component (i.e., $\varrho_{out}^i=\sqrt{E_i}\rho\sqrt{E_i}$) of the final state
%\begin{align*}
% \varrho_{out}^i=\frac{n_i}{2n}\Big[(1+s_3^i)\mathbb{I}+\frac{1-\sqrt{1-|s_i|^2}}{|s_i|^2}(s_1^is_3^i\sigma_1+s_2^is_3^i\sigma_2)\\
% +\Big(\sqrt{1-|s_i|^2}+\frac{1-\sqrt{1-|s_i|^2}}{|s_i|^2}s_3^i\Big)\sigma_3+\vec{s}_i.\vec{\sigma}\Big].
%\end{align*}
%The final state will have quantum coherence given by 
%\begin{align}\label{nelementpovmcoherence}
%	C_{l_1}(\delta_0')&=\sum_{i=1}^n\left|(s_{i1}-is_{i2})\Big(\frac{1}{2}+\frac{s_{i3}\big(n_i-\sqrt{n_i^2-n^2|s_i|^2}\big)}{2n|s_i|^2}\Big)\right|\nonumber\\
%	&+\sum_{i=1}^n\left|(s_{i1}+is_{i2})\Big(\frac{1}{2}+\frac{s_{i3}\big(n_i-\sqrt{n_i^2-n^2|s_i|^2}\big)}{2n|s_i|^2}\Big)\right|
%\end{align}
%The equation tells us that the contribution of raw quantumness from individual POVM to the final coherence is proportional to $\frac{1}{n}$. Hence the claim.
%$\square$

\subsection{Observations}
The analytical results and numerical simulations indicates that a measurement process may not always destroy coherence in the target state. It may create coherence also. This creation of coherence is the effect of non-zero off-diagonal terms present in the measurement elements, which we neoterize as `raw quantumness' in a measurement \footnote{There exist a lot of literatures which deals with the resource theory of operations, namely, Refs.\cite{PhysRevA.67.052301,PhysRevLett.122.190405,PhysRevA.95.062327,DATTA2018243}}. To show how this happens, first we consider POVMs with vanishing off-diagonal elements in Eq.(\ref{wer}), i.e., with $a_1=a_2=0$. For such a POVM the final Coherence created in an arbitrary qubit state (see Eq.{\ref{eqw}}) is 
\begin{align*}
	C_{l_1}(\rho'_{\pm})&=\beta \left|R\right| =\sqrt{r_1^2+r_2^2}(\sqrt{a_+^2-a_3^2}+\sqrt{a_-^2-a_3^2})\\
	&\leq \sqrt{r_1^2+r_2^2}\sqrt{1-4a_3^2}\leq\sqrt{r_1^2+r_2^2}.
\end{align*} 
Hence, such a measurement can not induce any coherence in an arbitrary qubit state. Similarly, we also consider an $n$-outcome POVM measurement from Eq. \ref{noutcomepovm}, with no off-diagonal terms, i.e., with $s_{i1}=s_{i2}=0$ $\forall$ $E_i's$. We consider its action on an arbitrary state $\rho=\frac{1}{2}(\mathbb{I}+\vec{r}.\vec{\sigma})$, so that the final coherence is given by
\begin{align*}
	C_{l_1}^n(\rho')&=\sqrt{r_1^2+r_2^2}\sum_{i}^n\sqrt{a_i^2-s_{i3}^2}\\
	&\leq\sqrt{r_1^2+r_2^2}\sum_{i}^na_i = \sqrt{r_1^2+r_2^2}.
\end{align*}
Motivated by these results, we can give the following definition

{\em Definition}.--Any qubit measurement with only diagonal POVM elements are free measurements, i.e., $E_i=\sum_k e_{kk}^i\ketbra{k}{k}$ with $\sum_ke_{kk}^i\leq 1$.

The qubit measurements with at least one non-diagonal POVM is not a free measurement and has potential to induce coherence in the target state. These measurements possess raw quantumness. The raw quantumness can be quantified as the `sum of absolute value of off-diagonal terms' in the measurements elements. We notice that the POVM elements in Eq. (\ref{wer}) have `raw quantumness' $C_{raw}=2\sqrt{a_1^2+a_2^2}$. Similarly, the raw quantumness in the POVM elements in Eq. (\ref{noutcomepovm}) is given as $C_{raw}=2a_i\sqrt{s_{i1}^2+s_{i2}^2}$. Interestingly, it can be seen from Eq. (\ref{sdf}) that the coherence induced in an incoherent state $\delta_0$ due to generalized POVM is proportional to $C_{raw}$. We also notice in Eq.(\ref{eqw}) that the coherence of final state is a function of the $C_{raw}$.
These leads us to the following observations.

For an arbitrary POVM measurement acting on qubits, the following statements are true
\begin{enumerate}
 \item If the initial state is incoherent, the induced coherence by a general two outcome POVM measurement in the final state is bounded as $0\leq C_{l_1}(\rho')\leq0.5$.
\item For initial incoherent state, if $C_{raw}=0$, the post measurement state will have no coherence.
 \item An initial incoherent and coherent state may acquire extra amount of coherence only if the measurement elements have non-zero raw quantumness.
\end{enumerate}
The above observations tell us that if the measurement under consideration is `quantum' enough then the decoherence due to measurement can be avoided. This will provide advantages in many quantum information processing tasks where measurement is a key element. Next, we provide an important Lemma for a measurement to have ability to create coherence.

\textit{Lemma-1}: The existence of raw quantumness in the POVM elements is a necessary but not sufficient condition to create coherence in qubits.

Necessary condition we have already proven. To show that it is not sufficient we present a counter example for which  even if $C_{raw}>0$, an incoherent state remains incoherent under the measurement. Consider the following $3$-outcome POVM measurement, 
\begin{equation}
 E_i=a_i(\mathbb{I}+\vec{r}_i.\vec{\sigma}), 
 \label{CNM-nIC}
\end{equation}
where $a_1=\frac{t}{3}$, $a_2=a_3=\frac{1}{2}(1-\frac{t}{3})$,  $\vec{r}_1=\{1,0,0\}^T$,  $\vec{r}_2=\{-b,\sqrt{1-b^2},0\}^T$, and $\vec{r}_3=\{-b,-\sqrt{1-b^2},0\}^T$, 
where $b=\frac{t}{3-t}$ and $t\in (0,1)$ \cite{2005JPhA...38.5979M}. This POVM  can not create coherence in an incoherent state.

\subsubsection{Coherence creation under application of successive POVM operators}
\begin{figure}[htb]
\centering
\includegraphics[scale=0.6]{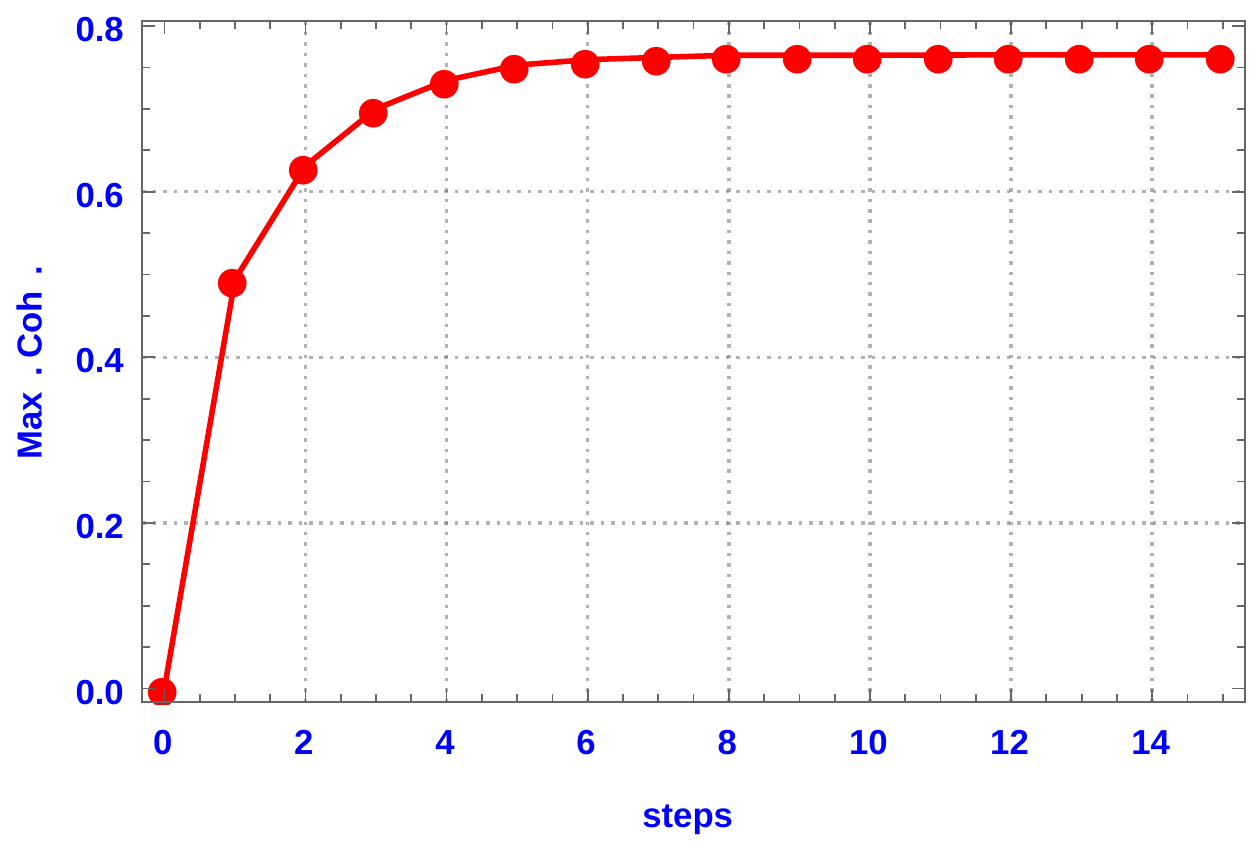}
\caption{Maximum coherence gain after application of $2$-outcome randomly generated POVM measurement on incoherent state $\delta_0$ in consecutive `steps'. To achieve the numerical maximum, we have created $2.2\times 10^5$ of random POVM using QETLAB for each `steps'. The plot shows that the maximum gain of coherence due to the consecutive application of POVM saturates at $0.76525$.}
\label{fig:ew}
\end{figure}

We know that, one can create non-zero coherence from an incoherent state if we apply general measurement. This observation prompts us to investigate how much coherence one can create from an incoherent state if one allows to perform the measurement consecutively many times (`steps'). For our analysis, we consider a qubit incoherent state $\delta_0$ and two-outcome POVM. We create random $2$-outcome POVM ($\approx 2.2\times 10^5$) to obtain the target state with maximum coherence for each step. Our numerical simulation has been plotted in the Fig.\ref{fig:ew}. The Fig.\ref{fig:ew} shows that the maximum coherence one can reach from the qubit incoherent state $\delta_0$ is $0.76525$ unit.
%%%%%%%%%%%%%%%%%%%%%%%%%%%%%%%%%%%%%%%%%%%%%%%%%%%%%%%%%%%%%%%5
\section{Two categories of measurements based on coherence resource theory}
We may consider measurement, $M$, as channel, i.e., for set of POVMs $\{E_i,\forall i\in I^+\}$ and the state $\rho\in\mathbb{C}_d$,
\begin{equation}
\label{eqty}
\Lambda_M(\rho)=\sum_i\sqrt{E_i}\rho\sqrt{E_i},
\end{equation}
where the form of the Kraus operators as $K_i=U_i\sqrt{E_i}$ with $U_i$ being arbitrary unitary. Note that the choice $E_i=K_i^{\dagger}K_i$ is not unique. The above channel is {\em unital} as $\Lambda_M(\mathbb{I})=\mathbb{I}$. Based on above findings, we will categorize the non-selective measurements in two categories -- measurements which do not create coherence and which do. The formal definition for such nomenclature is given below:\\
{\em Definition}.-- A measurement which does not create coherence is defined by $\Lambda_M(\mathcal{I})\subset\mathcal{I}$, where $\mathcal{I}$ is the set of all incoherent states.

Therefore the measurements which violate $\Lambda_M(\mathcal{I})\subset\mathcal{I}$, will create coherence in the state. Some properties of `coherence non-generating measurement (CNM)' can readily be listed below
\begin{enumerate}
 \item IC $\subset$ CNM $\subset$ CNC, where IC is the set of incoherent operations \cite{Streltsov_Coh} and CNC are the set of coherence-non-generating channels \cite{PhysRevA.94.012326}.
 \item If two measurements $M_1$ and $M_2$ are CNM, then the composition $\Lambda_{M_1}\circ \Lambda_{M_2}$ as well as tensor product $\Lambda_{M_1}\otimes \Lambda_{M_2}$ are CNM.
 \item The $l_1$-norm of coherence for qubits, and the relative entropy of coherence for arbitrary dimension never increases under CNM.
\end{enumerate}

\begin{proof}
 From the Ref.\cite{PhysRevA.94.012326}, we know that IC $\subset$ CNC. Now, we know that any unital channels can be transformed into a measurement channels by choosing some appropriate unitaries \cite{PhysRevA.71.012330,Paris2012}. This fact easily led us to conclude that CNM $\subset$ CNC. Now as there may exist some CNMs which are not incoherent as its `modified Kraus elements' are not individually incoherent (see Eq.\ref{CNM-nIC}). Hence, the relation IC $\subset$ CNM $\subset$ CNC.

 The composition of two measurement channels can be defined as $\Lambda_{M_1}\circ \Lambda_{M_2}(\rho)=\sum_j\sqrt{E_j^1}\big(\sum_i\sqrt{E_i^2}\rho\sqrt{E_i^2}\big)\sqrt{E_j^1}$. Then one can readily prove that the composition of two CNMs is also a CNM. The tensor product of two measurements is defined in the bipartite systems, $\rho_{AB}$, i.e., $\Lambda_{M_1}\otimes \Lambda_{M_2}(\rho_{AB})$. As any bipartite incoherent state can be written as product of two incoherent state in their fixed local basis, one readily prove that tensor product of two CNMs is a CNM. 
 
 As for any qubit state, $\rho$, $C_{l_1}(\rho)=C_{\rm tr}(\rho)$ \cite{PhysRevA.93.012110}, we find that 
 \begin{align*}
  C_{l_1}(\Lambda_{CNM}(\rho))= ||\Lambda_{CNM}(\rho)-\Lambda_{CNM}(\rho)^D||_{\rm tr},\\
  \leq ||\Lambda_{CNM}(\rho)-\Lambda_{CNM}(\rho^D)||_{\rm tr},\\
  \leq ||\rho-\rho^D||_{\rm tr}=C_{l_1}(\rho),
 \end{align*}
where the first inequality is because $\Lambda_{CNM}(\rho^D)$ may not be the closest incoherent state to $\Lambda_{CNM}(\rho)$ and the second inequality is from the contractive nature of trace distance under CPTP map. However, for higher dimensional system this property may not hold.

From the monotonicity property of relative entropy under CPTP map, one can find that 
\begin{align*}
 C_r(\rho)=S(\rho||\rho^D)\geq S(\Lambda_{CNM}(\rho)||\Lambda_{CNM}(\rho^D))\\
 \geq \min_{\delta\in\mathcal{I}}S(\Lambda_{CNM}(\rho)||\delta)=C_r(\Lambda_{CNM}(\rho)).
\end{align*}
Note that in the above proof, we use the fact that the state $\Lambda_{CNM}(\rho^D)$ is still incoherent but may not be the optimal one.
\end{proof}
This classification of measurement might give us new insight to the quantum coherence theory. The following important remarks can be made 
\begin{enumerate}
 \item In coherence distillation, CNM does not give extra advantage as it is CNM $\subset$ CNC \cite{PhysRevA.94.012326}.
 \item As for qubits coherence of formation $C_f(\rho)$ is monotonic function of $C_{l_1}(\rho)$ \cite{PhysRevA.94.012326}, therefore $\forall \Lambda_{CNM}$, $C_f(\Lambda_{CNM}(\rho))\leq C_f(\rho)$. However, for multiqubit states as well as for higher dimensional states this may not hold.
 \item There exists some CNMs which are not always IC. 
\end{enumerate}
To illustrate the above remarks we consider the following CNM measurements:

{\em Example}.-- The action of the tensor product of the identity measurement and the POVM measurement with elements  
\begin{align*}
E_1=
  \left[ {\begin{array}{cc}
   \frac{1}{2} & -\frac{1}{2\sqrt{2}} \\
   -\frac{1}{2\sqrt{2}} & \frac{1}{2} \\
  \end{array} } \right]
  \:\:\mbox{and}\:\:
 E_2= \left[ {\begin{array}{cc}
   \frac{1}{2} & \frac{1}{2\sqrt{2}} \\
   \frac{1}{2\sqrt{2}} & \frac{1}{2} \\
  \end{array} } \right]
\end{align*}
(which are both CNM), on the maximally entangled qubit state $\ket{\psi_+}=\frac{1}{\sqrt{2}}(\ket{00}+\ket{11})$, 
lead to the state $\rho_f=\frac{1}{2}(\ketbra{e_1}{e_1}+\ketbra{e_2}{e_2})$, where 
$\ket{e_1}=\frac{1}{4}(h_+\ket{00}+h_-(\ket{01}+\ket{10})+h_+\ket{11})$ and 
$\ket{e_2}=\frac{1}{4}(h_+\ket{00}-h_-(\ket{01}+\ket{10})+h_+\ket{11})$ with 
$h_{\pm}=\sqrt{2-\sqrt{2}}\pm\sqrt{2-\sqrt{2}}$. We know that $C_f(\ket{\psi_+})=1$. Now to calculate the coherence of 
formation of $\rho_f$, we notice that $\rho_f$ lives in the subspace spanned by vectors $\ket{e_1}$ and $\ket{e_2}$. 
Therefore, pure state decomposition of $\rho_f=\sum_ip_i\rho_i$, where  $\rho_i=\ketbra{\psi_i}{\psi_i}$ 
with $\ket{\psi_i}=\cos\theta\ket{e_1}+\sin\theta e^{\mi\phi}\ket{e_2}$, $\theta\in [0,\frac{\pi}{2}]$ and 
$\phi\in [0,2\pi]$. Hence, one can readily calculate and find that 
$\forall \rho_i$, $S(\rho_i^D)=-\epsilon_{+}\log\epsilon_+-\epsilon_{-}\log\epsilon_-=1+H_2(\epsilon_+)>1$ holds, where 
$\epsilon_{\pm}=\frac{1}{4}(2\pm\sqrt{
2})|\cos\theta\pm e^{\mi\phi}\sin\theta|^2$. Therefore, the $C_f(\rho_f)=\min_{\{p_i,\ket{\psi_i}\}}\sum_ip_iS(\rho_i^D)\geq \min_{\{\ket{\psi_i}\}} S(\rho_i^D)>1$. This proves that the coherence of formation may increase under CNMs.

Above example and the example in Eq.(\ref{CNM-nIC}) show that although the measurement is CNM, it is not IC as individual POVM elements may induce coherence in the state.

%%%%%%%%%%%%%%%%%%%%%%%%%%%%%%%%%%%%%%%%%%%%%%%%%%%%%%%%%%%%%%%%%%%%
\section{Coherence creation under POVM measurement and entanglement}
\label{new-coh-ent}
Recently, it was shown that for bipartite quantum systems, if one performs von Neumann measurement on one half of the system by bringing an apparatus, then the induced distillable entanglement\footnote{It quantifies the \# of Bell states one can extract from $N$ number of mixed entangled states. For further study, we refer readers to \cite{PhysRevA.53.2046}.} between system and apparatus bipartition is exactly equal to the one-way work deficit \footnote{The work deficit is the information, or work, that cannot be extracted from a bipartite quantum state when the parties are in distant locations, as compared to the case when the same are together. For more information on work deficit, we direct the enthusiastic reader to Refs. \cite{PhysRevLett.89.180402,PhysRevLett.90.100402,PhysRevA.71.062307}} present in the initial bipartite system \cite{Link-Disc,PhysRevLett.106.220403}. This interesting piece of result relates quantum correlations beyond entanglement with the distillable entanglement.
	
Here, in this work, we will connect two important resources, namely, the entanglement and the quantum coherence through POVM measurements. Let us consider a bipartite product state, $\rho_1=\rho_A\otimes\rho_S$, where $A$ denotes the state for ancilla and $S$ for system. Now, from Neumark's dilation theorem, we know that the POVM $\{E_i\}$ on the system is equivalent to unitary evolution and the projective measurement on the ancilla, i.e., 
\begin{align}
\rho_1\mapsto \rho_2^M=\sum_i(\Pi_i\otimes\mathbb{I})\rho_{AS}(\Pi_i\otimes\mathbb{I}),
\end{align}
where $\rho_{AS}=U (\rho_A\otimes\rho_S)U^{\dagger}$ with $U$ being global unitary and the projective measurement is being done with apparatus $M$. It was shown in \cite{Link-Disc}, that using this method we can create distillable entanglement between the apparatus and the state $\rho_{AS}$ given as $E^{M|AS}(\rho_2^M)=S(\sum_i\Pi_i\otimes\mathbb{I}\rho_{AS}\Pi_i\otimes\mathbb{I})-S(\rho_{AS})$. Creation of entanglement is possible only if $\rho_{AS}$ has non-zero discord in it \cite{Link-Disc}.
%The evolve state after this interaction is $\rho_{AS}=U (\rho_A\otimes\rho_S)U^{\dagger}$. 
However, we are interested in the following quantity,
\begin{align}\label{disent}
   E_{min}^{M|AS}(\rho_2^M)=\min_{\Pi_i}S\left(\sum_i\Pi_i\otimes\mathbb{I}\rho_{AS}\Pi_i\otimes\mathbb{I}\right)-S(\rho_{AS}), 
\end{align}
which is the entanglement obtained by minimizing over the projective measurements. The minimum entanglement created by this method is also known as the one way information deficit : $E_{min}^{M|AS}(\rho_2^M)=\overrightarrow{\triangle}(\rho_{AS})$ \cite{PhysRevLett.89.180402,PhysRevA.71.062307}.

By definition we have $S(\rho_2^M)-S(\rho_{AS})\geq E_{min}^{M|AS}(\rho_2^M)$. Using the fact that $S(\rho_{AS})=S(\rho_A)+S(\rho_S)$ and $S(\rho_2^M)\leq S(\rho_S^M)+S(\rho_A^M)$, where $\rho_S^M={\rm Tr}_A[\rho_2^M]$, we reach to
\begin{align}
    E_{min}^{M|AS}(\rho_2^M)&\leq S(\rho_A^D)+S(\rho_S^M)-S(\rho_A)-S(\rho_S)\nonumber\\
   & =C_R(\rho_A)+S(\rho_S^M)-S(\rho_S).
\end{align}
Since the projective measurement is done on the apparatus state $\rho_A$, $\rho_A^M=\rho_A^D$ in the basis of projection. Now, if we concentrate on the quantity, $S(\rho_S^M)-S(\rho_S)$, we find that it can be written as 
\begin{align*}
    S(\rho_S^M)-S(\rho_S)=S(\rho_S^{MD})-S(\rho_S)-C_R(\rho_S^M),
\end{align*}
where $C_R(\rho_S^M)$ is the coherence of $\rho_S$ after POVM is performed on the system $S$, i.e., $C_R(\rho_S^M)=C_R(\rho_S^{POVM})$, where $\rho_S^{MD}$ is diagonal version of $\rho_S^M$. Therefore, we finally find that
\begin{align}
  &E_{min}^{M|AS}(\rho_2^M)+C_R(\rho_S^{ POVM})+S(\rho_S)\leq C_R(\rho_A)+S(\rho_S^{MD}),\nonumber\\
  &E_{min}^{M|AS}(\rho_2^M)+C_R(\rho_S^{ POVM})+S(\rho_S) \leq \log NM,
  \label{ent_coh_mix}
\end{align}
where $N$ and $M$ are the dimension of the system $S$ and apparatus $A$ respectively and $S(\rho_S)$ denotes mixedness of the system $\rho_S$. The inequality in Eq.(\ref{ent_coh_mix}) relates two important resources, namely, the minimum entanglement and the coherence through the POVM measurement. It implies that the coherence created will be less if the initial system is highly mixed and/or the minimum entanglement that can be created is large. This inequality is similar to the complementarity relation between Coherence and entanglement for an arbitrary bipartite state \cite{Xi2015}. 
	%%%%%%%%%%%%%%%%%%%%%%%%%%%%%%%%%%%%%%%%%%%%%%%%%%%%%%%%%%%%5555
\section{Conclusion}
  \label{Conclusion}  
The emergent quantum technologies exploit the resource available in two main ingredients – quantum states and the allowed quantum operations \cite{PhysRevA.67.052301,PhysRevLett.122.190405}. Therefore, it is important to study the properties of quantum operations, mainly, its resource creating ability. While many operations can create coherence in the physical system, specifically, how much coherence can be created by a non-selective general measurement process was not explored in detail earlier. We have studied this phenomenon for both projective and generalized POVM operations on qubits. We find the maximum amount coherence that can be induced in an incoherent qubit state. This maximum coherence is 1/2 for a pure incoherent state. It is also demonstrated that it is not always possible to increase the coherence of a qubit state although it is still possible to prevent loss of coherence.

Interestingly, we find that the more elements present in the POVM sets for a measurement, the less is its coherence creation ability. Specifically, our result indicates that the maximum coherence generated from an incoherent state, using n outcome random POVM decreases exponentially with $n$. Since more POVM elements mean, in the dilated Hilbert space, we have a high-dimensional ancillary system. This shows that when a quantum system interacts with a larger system, the ability to create coherence decreases exponentially. Furthermore, to explain the ability of coherence creation by the non-selective general measurement process, we introduce the notion of ‘raw’ quantumness in POVM elements. We show that ‘raw quantumness’ in the POVM elements is a necessary but not sufficient condition in order to induce coherence in a qubit state. 

Lastly, we have also demonstrated the creation of coherence in a bipartite state and show that it is dependent on the mixedness of the subsystem and the entanglemnt developed between apparatus and the bipartite system. We believe these findings through new lights on the role of measurement in the creation of quantum coherence.

As a future line of work, it would be interesting to obtain the analytical results for two outcome and n-outcome POVMs. Also important is to obtain, how the coherence creation capability decreases with increasing $n$. Last but not the least, it will be useful to extend this work on higher dimensional states.

{\em Acknowledgements}.-- Sk Sazim would like to acknowledge the financial support from AK Pati's JC Bose Fellowship Grant
during 14 July--20 Sept, 2019 at HRI. He also acknowledges the financial support through the {\v S}tefan Schwarz stipend from Slovak Academy of Sciences, Bratislava, OPTIQUTE (APVV-18-0518) and HOQIP (VEGA 2/0161/19). 
Sanuja Mohanty would like to thank HRI for the financial support during her visits in 2017 and 2018.


\begin{thebibliography}{40}%
\makeatletter
\providecommand \@ifxundefined [1]{%
 \@ifx{#1\undefined}
}%
\providecommand \@ifnum [1]{%
 \ifnum #1\expandafter \@firstoftwo
 \else \expandafter \@secondoftwo
 \fi
}%
\providecommand \@ifx [1]{%
 \ifx #1\expandafter \@firstoftwo
 \else \expandafter \@secondoftwo
 \fi
}%
\providecommand \natexlab [1]{#1}%
\providecommand \enquote  [1]{``#1''}%
\providecommand \bibnamefont  [1]{#1}%
\providecommand \bibfnamefont [1]{#1}%
\providecommand \citenamefont [1]{#1}%
\providecommand \href@noop [0]{\@secondoftwo}%
\providecommand \href [0]{\begingroup \@sanitize@url \@href}%
\providecommand \@href[1]{\@@startlink{#1}\@@href}%
\providecommand \@@href[1]{\endgroup#1\@@endlink}%
\providecommand \@sanitize@url [0]{\catcode `\\12\catcode `\$12\catcode
  `\&12\catcode `\#12\catcode `\^12\catcode `\_12\catcode `\%12\relax}%
\providecommand \@@startlink[1]{}%
\providecommand \@@endlink[0]{}%
\providecommand \url  [0]{\begingroup\@sanitize@url \@url }%
\providecommand \@url [1]{\endgroup\@href {#1}{\urlprefix }}%
\providecommand \urlprefix  [0]{URL }%
\providecommand \Eprint [0]{\href }%
\providecommand \doibase [0]{http://dx.doi.org/}%
\providecommand \selectlanguage [0]{\@gobble}%
\providecommand \bibinfo  [0]{\@secondoftwo}%
\providecommand \bibfield  [0]{\@secondoftwo}%
\providecommand \translation [1]{[#1]}%
\providecommand \BibitemOpen [0]{}%
\providecommand \bibitemStop [0]{}%
\providecommand \bibitemNoStop [0]{.\EOS\space}%
\providecommand \EOS [0]{\spacefactor3000\relax}%
\providecommand \BibitemShut  [1]{\csname bibitem#1\endcsname}%
\let\auto@bib@innerbib\@empty
%</preamble>
\bibitem [{\citenamefont {{Aberg}}(2006)}]{J_Aberg_Coh}%
  \BibitemOpen
  \bibfield  {author} {\bibinfo {author} {\bibfnamefont {J.}~\bibnamefont
  {{Aberg}}},\ }\href@noop {} {\bibfield  {journal} {\bibinfo  {journal}
  {eprint arXiv:quant-ph/0612146}\ } (\bibinfo {year} {2006})},\ \Eprint
  {http://arxiv.org/abs/quant-ph/0612146} {quant-ph/0612146} \BibitemShut
  {NoStop}%
\bibitem [{\citenamefont {Baumgratz}\ \emph {et~al.}(2014)\citenamefont
  {Baumgratz}, \citenamefont {Cramer},\ and\ \citenamefont
  {Plenio}}]{Baumg_Cohere}%
  \BibitemOpen
  \bibfield  {author} {\bibinfo {author} {\bibfnamefont {T.}~\bibnamefont
  {Baumgratz}}, \bibinfo {author} {\bibfnamefont {M.}~\bibnamefont {Cramer}}, \
  and\ \bibinfo {author} {\bibfnamefont {M.~B.}\ \bibnamefont {Plenio}},\
  }\href {\doibase 10.1103/PhysRevLett.113.140401} {\bibfield  {journal}
  {\bibinfo  {journal} {Phys. Rev. Lett.}\ }\textbf {\bibinfo {volume} {113}},\
  \bibinfo {pages} {140401} (\bibinfo {year} {2014})}\BibitemShut {NoStop}%
\bibitem [{\citenamefont {Winter}\ and\ \citenamefont
  {Yang}(2016)}]{Winter_Cohere}%
  \BibitemOpen
  \bibfield  {author} {\bibinfo {author} {\bibfnamefont {A.}~\bibnamefont
  {Winter}}\ and\ \bibinfo {author} {\bibfnamefont {D.}~\bibnamefont {Yang}},\
  }\href {\doibase 10.1103/PhysRevLett.116.120404} {\bibfield  {journal}
  {\bibinfo  {journal} {Phys. Rev. Lett.}\ }\textbf {\bibinfo {volume} {116}},\
  \bibinfo {pages} {120404} (\bibinfo {year} {2016})}\BibitemShut {NoStop}%
\bibitem [{\citenamefont {Streltsov}\ \emph {et~al.}(2017)\citenamefont
  {Streltsov}, \citenamefont {Adesso},\ and\ \citenamefont
  {Plenio}}]{Streltsov_Coh}%
  \BibitemOpen
  \bibfield  {author} {\bibinfo {author} {\bibfnamefont {A.}~\bibnamefont
  {Streltsov}}, \bibinfo {author} {\bibfnamefont {G.}~\bibnamefont {Adesso}}, \
  and\ \bibinfo {author} {\bibfnamefont {M.~B.}\ \bibnamefont {Plenio}},\
  }\href {\doibase 10.1103/RevModPhys.89.041003} {\bibfield  {journal}
  {\bibinfo  {journal} {Rev. Mod. Phys.}\ }\textbf {\bibinfo {volume} {89}},\
  \bibinfo {pages} {041003} (\bibinfo {year} {2017})}\BibitemShut {NoStop}%
\bibitem [{\citenamefont {Wu}\ \emph {et~al.}(2020)\citenamefont
  {Wu}, \citenamefont {Plenio},\ and\ \citenamefont
  {Streltsov}}]{Wu2020}%
  \BibitemOpen
  \bibfield  {author} {\bibinfo {author} {\bibfnamefont {K.-D.}~\bibnamefont
  {Wu}}, \bibinfo {author} {\bibfnamefont {T.}~\bibnamefont {Theurer}}, \bibinfo {author} {\bibfnamefont {C.-F.}~\bibnamefont {Li}}, \bibinfo {author} {\bibfnamefont {G.-C.}~\bibnamefont {Guo}}, \bibinfo {author} {\bibfnamefont {M.~B.}\ \bibnamefont {Plenio}},\
  and\ \bibinfo {author} {\bibfnamefont {A.}\ \bibnamefont {Streltsov}},\
  }\href {\doibase 10.1038/s41534-020-0250-z} {\bibfield  {journal}
  {\bibinfo  {journal} {npj Quantum Information}\ }\textbf {\bibinfo {volume} {6}},\
  \bibinfo {pages} {22} (\bibinfo {year} {2020})}\BibitemShut {NoStop}% 
\bibitem [{\citenamefont {Bandyopadhyay}(2020)}]{PhysRevA.102.050202}%
  \BibitemOpen
  \bibfield  {author} {\bibinfo {author} {\bibfnamefont {S.}~\bibnamefont
  {Bandyopadhyay}},\ }\href {\doibase 10.1103/PhysRevA.102.050202} {\bibfield  {journal}
  {\bibinfo  {journal} {Phys. Rev. A}\ }\textbf {\bibinfo {volume} {102}},\
  \bibinfo {pages} {050202} (\bibinfo {year} {2020})}\BibitemShut {NoStop}%  
\bibitem [{\citenamefont {{Kiukas}}\ \emph {et~al.}(2020)\citenamefont
  {Kiukas}, \citenamefont {McNulty},\ and\ \citenamefont
  {Pellonp{\"a}{\"a}}}]{2020arXiv201107239K}%
  \BibitemOpen
  \bibfield  {author} {\bibinfo {author} {\bibfnamefont {J.}~\bibnamefont
  {{Kiukas}}}, \bibinfo {author} {\bibfnamefont {D.}~\bibnamefont {McNulty}}, \
  and\ \bibinfo {author} {\bibfnamefont {J.-P.}\ \bibnamefont {Pellonp{\"a}{\"a}}},\ }\href@noop {} {\bibfield  {journal} {\bibinfo  {journal}
  {eprint arXiv:2011.07239}\ } (\bibinfo {year} {2020})},\ \Eprint
  {http://arxiv.org/abs/quant-ph:2011.07239} {} \BibitemShut
  {NoStop}%  
\bibitem [{\citenamefont {Bischof}\ \emph {et~al.}(2021)\citenamefont
  {Bischof}, \citenamefont {Kampermann},\ and\ \citenamefont
  {Bru\ss{}}}]{PhysRevA.103.032429}%
  \BibitemOpen
  \bibfield  {author} {\bibinfo {author} {\bibfnamefont {F.}~\bibnamefont
  {Bischof}}, \bibinfo {author} {\bibfnamefont {H.}~\bibnamefont {Kampermann}}, \
  and\ \bibinfo {author} {\bibfnamefont {D.}\ \bibnamefont {Bru\ss{}}},\
  }\href {\doibase 10.1103/PhysRevA.103.032429} {\bibfield  {journal}
  {\bibinfo  {journal} {Phys. Rev. A}\ }\textbf {\bibinfo {volume} {103}},\
  \bibinfo {pages} {032429} (\bibinfo {year} {2021})}\BibitemShut {NoStop}%
\bibitem [{\citenamefont {Mani}\ and\ \citenamefont
  {Karimipour}(2015)}]{Mani_CP}%
  \BibitemOpen
  \bibfield  {author} {\bibinfo {author} {\bibfnamefont {A.}~\bibnamefont
  {Mani}}\ and\ \bibinfo {author} {\bibfnamefont {V.}~\bibnamefont
  {Karimipour}},\ }\href {\doibase 10.1103/PhysRevA.92.032331} {\bibfield
  {journal} {\bibinfo  {journal} {Phys. Rev. A}\ }\textbf {\bibinfo {volume}
  {92}},\ \bibinfo {pages} {032331} (\bibinfo {year} {2015})}\BibitemShut
  {NoStop}%
\bibitem [{\citenamefont {{Garc{\'\i}a-D{\'\i}az}}\ \emph
  {et~al.}(2016)\citenamefont {{Garc{\'\i}a-D{\'\i}az}}, \citenamefont
  {{Egloff}},\ and\ \citenamefont {{Plenio}}}]{2015arXiv151006683G}%
  \BibitemOpen
  \bibfield  {author} {\bibinfo {author} {\bibfnamefont {M.}~\bibnamefont
  {{Garc{\'\i}a-D{\'\i}az}}}, \bibinfo {author} {\bibfnamefont
  {D.}~\bibnamefont {{Egloff}}}, \ and\ \bibinfo {author} {\bibfnamefont
  {M.~B.}\ \bibnamefont {{Plenio}}},\ }\href@noop {} {\bibfield  {journal}
  {\bibinfo  {journal} {Quant. Inf. Comp.}\ }\textbf {\bibinfo {volume} {16}},\
  \bibinfo {pages} {1282} (\bibinfo {year} {2016})},\ \Eprint
  {http://arxiv.org/abs/1510.06683} {arXiv:1510.06683 [quant-ph]} \BibitemShut
  {NoStop}%
\bibitem [{\citenamefont {Zanardi}\ \emph
  {et~al.}(2017{\natexlab{a}})\citenamefont {Zanardi}, \citenamefont
  {Styliaris},\ and\ \citenamefont {Campos~Venuti}}]{PhysRevA.95.052306}%
  \BibitemOpen
  \bibfield  {author} {\bibinfo {author} {\bibfnamefont {P.}~\bibnamefont
  {Zanardi}}, \bibinfo {author} {\bibfnamefont {G.}~\bibnamefont {Styliaris}},
  \ and\ \bibinfo {author} {\bibfnamefont {L.}~\bibnamefont {Campos~Venuti}},\
  }\href {\doibase 10.1103/PhysRevA.95.052306} {\bibfield  {journal} {\bibinfo
  {journal} {Phys. Rev. A}\ }\textbf {\bibinfo {volume} {95}},\ \bibinfo
  {pages} {052306} (\bibinfo {year} {2017}{\natexlab{a}})}\BibitemShut
  {NoStop}%
\bibitem [{\citenamefont {Zanardi}\ \emph
  {et~al.}(2017{\natexlab{b}})\citenamefont {Zanardi}, \citenamefont
  {Styliaris},\ and\ \citenamefont {Campos~Venuti}}]{PhysRevA.95.052307}%
  \BibitemOpen
  \bibfield  {author} {\bibinfo {author} {\bibfnamefont {P.}~\bibnamefont
  {Zanardi}}, \bibinfo {author} {\bibfnamefont {G.}~\bibnamefont {Styliaris}},
  \ and\ \bibinfo {author} {\bibfnamefont {L.}~\bibnamefont {Campos~Venuti}},\
  }\href {\doibase 10.1103/PhysRevA.95.052307} {\bibfield  {journal} {\bibinfo
  {journal} {Phys. Rev. A}\ }\textbf {\bibinfo {volume} {95}},\ \bibinfo
  {pages} {052307} (\bibinfo {year} {2017}{\natexlab{b}})}\BibitemShut
  {NoStop}%
\bibitem [{\citenamefont {Bu}\ \emph {et~al.}(2017{\natexlab{a}})\citenamefont
  {Bu}, \citenamefont {Kumar}, \citenamefont {Zhang},\ and\ \citenamefont
  {Wu}}]{Bu_CP}%
  \BibitemOpen
  \bibfield  {author} {\bibinfo {author} {\bibfnamefont {K.}~\bibnamefont
  {Bu}}, \bibinfo {author} {\bibfnamefont {A.}~\bibnamefont {Kumar}}, \bibinfo
  {author} {\bibfnamefont {L.}~\bibnamefont {Zhang}}, \ and\ \bibinfo {author}
  {\bibfnamefont {J.}~\bibnamefont {Wu}},\ }\href {\doibase
  https://doi.org/10.1016/j.physleta.2017.03.022} {\bibfield  {journal}
  {\bibinfo  {journal} {Physics Letters A}\ }\textbf {\bibinfo {volume}
  {381}},\ \bibinfo {pages} {1670 } (\bibinfo {year}
  {2017}{\natexlab{a}})}\BibitemShut {NoStop}%
\bibitem [{\citenamefont {Styliaris}\ \emph {et~al.}(2018)\citenamefont
  {Styliaris}, \citenamefont {Campos~Venuti},\ and\ \citenamefont
  {Zanardi}}]{PhysRevA.97.032304}%
  \BibitemOpen
  \bibfield  {author} {\bibinfo {author} {\bibfnamefont {G.}~\bibnamefont
  {Styliaris}}, \bibinfo {author} {\bibfnamefont {L.}~\bibnamefont
  {Campos~Venuti}}, \ and\ \bibinfo {author} {\bibfnamefont {P.}~\bibnamefont
  {Zanardi}},\ }\href {\doibase 10.1103/PhysRevA.97.032304} {\bibfield
  {journal} {\bibinfo  {journal} {Phys. Rev. A}\ }\textbf {\bibinfo {volume}
  {97}},\ \bibinfo {pages} {032304} (\bibinfo {year} {2018})}\BibitemShut
  {NoStop}%
\bibitem [{\citenamefont {{Zhang}}\ \emph {et~al.}(2018)\citenamefont
  {{Zhang}}, \citenamefont {{Ma}}, \citenamefont {{Chen}},\ and\ \citenamefont
  {{Fei}}}]{2018QuIP...17..186Z}%
  \BibitemOpen
  \bibfield  {author} {\bibinfo {author} {\bibfnamefont {L.}~\bibnamefont
  {{Zhang}}}, \bibinfo {author} {\bibfnamefont {Z.}~\bibnamefont {{Ma}}},
  \bibinfo {author} {\bibfnamefont {Z.}~\bibnamefont {{Chen}}}, \ and\ \bibinfo
  {author} {\bibfnamefont {S.-M.}\ \bibnamefont {{Fei}}},\ }\href {\doibase
  10.1007/s11128-018-1928-4} {\bibfield  {journal} {\bibinfo  {journal}
  {Quantum Information Processing}\ }\textbf {\bibinfo {volume} {17}},\
  \bibinfo {eid} {186} (\bibinfo {year} {2018})},\ \Eprint
  {http://arxiv.org/abs/1711.02458} {arXiv:1711.02458 [quant-ph]} \BibitemShut
  {NoStop}%
\bibitem [{\citenamefont {Hu}(2016)}]{PhysRevA.94.012326}%
  \BibitemOpen
  \bibfield  {author} {\bibinfo {author} {\bibfnamefont {X.}~\bibnamefont
  {Hu}},\ }\href {\doibase 10.1103/PhysRevA.94.012326} {\bibfield  {journal}
  {\bibinfo  {journal} {Phys. Rev. A}\ }\textbf {\bibinfo {volume} {94}},\
  \bibinfo {pages} {012326} (\bibinfo {year} {2016})}\BibitemShut {NoStop}%
\bibitem [{\citenamefont {Nielsen}\ and\ \citenamefont
  {Chuang}(2010)}]{NielsonBook}%
  \BibitemOpen
  \bibfield  {author} {\bibinfo {author} {\bibfnamefont {M.~A.}\ \bibnamefont
  {Nielsen}}\ and\ \bibinfo {author} {\bibfnamefont {I.~L.}\ \bibnamefont
  {Chuang}},\ }\href {\doibase 10.1017/CBO9780511976667} {\emph {\bibinfo
  {title} {Quantum Computation and Quantum Information: 10th Anniversary
  Edition}}}\ (\bibinfo  {publisher} {Cambridge University Press},\ \bibinfo
  {year} {2010})\BibitemShut {NoStop}%
\bibitem [{\citenamefont {Singh}\ \emph {et~al.}(2016)\citenamefont {Singh},
  \citenamefont {Pati},\ and\ \citenamefont {Bera}}]{math4030047}%
  \BibitemOpen
  \bibfield  {author} {\bibinfo {author} {\bibfnamefont {U.}~\bibnamefont
  {Singh}}, \bibinfo {author} {\bibfnamefont {A.~K.}\ \bibnamefont {Pati}}, \
  and\ \bibinfo {author} {\bibfnamefont {M.~N.}\ \bibnamefont {Bera}},\ }\href
  {\doibase 10.3390/math4030047} {\bibfield  {journal} {\bibinfo  {journal}
  {Mathematics}\ }\textbf {\bibinfo {volume} {4}} (\bibinfo {year} {2016}),\
  10.3390/math4030047}\BibitemShut {NoStop}%
\bibitem [{\citenamefont {Cheng}\ and\ \citenamefont
  {Hall}(2015)}]{PhysRevA.92.042101}%
  \BibitemOpen
  \bibfield  {author} {\bibinfo {author} {\bibfnamefont {S.}~\bibnamefont
  {Cheng}}\ and\ \bibinfo {author} {\bibfnamefont {M.~J.~W.}\ \bibnamefont
  {Hall}},\ }\href {\doibase 10.1103/PhysRevA.92.042101} {\bibfield  {journal}
  {\bibinfo  {journal} {Phys. Rev. A}\ }\textbf {\bibinfo {volume} {92}},\
  \bibinfo {pages} {042101} (\bibinfo {year} {2015})}\BibitemShut {NoStop}%
\bibitem [{\citenamefont {Bu}\ \emph {et~al.}(2017{\natexlab{b}})\citenamefont
  {Bu}, \citenamefont {Singh}, \citenamefont {Fei}, \citenamefont {Pati},\ and\
  \citenamefont {Wu}}]{PhysRevLett.119.150405}%
  \BibitemOpen
  \bibfield  {author} {\bibinfo {author} {\bibfnamefont {K.}~\bibnamefont
  {Bu}}, \bibinfo {author} {\bibfnamefont {U.}~\bibnamefont {Singh}}, \bibinfo
  {author} {\bibfnamefont {S.-M.}\ \bibnamefont {Fei}}, \bibinfo {author}
  {\bibfnamefont {A.~K.}\ \bibnamefont {Pati}}, \ and\ \bibinfo {author}
  {\bibfnamefont {J.}~\bibnamefont {Wu}},\ }\href {\doibase
  10.1103/PhysRevLett.119.150405} {\bibfield  {journal} {\bibinfo  {journal}
  {Phys. Rev. Lett.}\ }\textbf {\bibinfo {volume} {119}},\ \bibinfo {pages}
  {150405} (\bibinfo {year} {2017}{\natexlab{b}})}\BibitemShut {NoStop}%
\bibitem [{\citenamefont {Neumann}(1955)}]{VonNeumann}%
  \BibitemOpen
  \bibfield  {author} {\bibinfo {author} {\bibfnamefont {J.~V.}\ \bibnamefont
  {Neumann}},\ }\href@noop {} {\emph {\bibinfo {title} {Mathematical
  Foundations of Quantum Mechanics}}}\ (\bibinfo  {publisher} {Princeton
  University Press},\ \bibinfo {address} {New York},\ \bibinfo {year}
  {1955})\BibitemShut {NoStop}%
\bibitem [{\citenamefont {Wheeler}\ and\ \citenamefont
  {Zurek}()}]{measurementWheeler}%
  \BibitemOpen
  \bibinfo {editor} {\bibfnamefont {J.~A.}\ \bibnamefont {Wheeler}}\ and\
  \bibinfo {editor} {\bibfnamefont {W.~H.}\ \bibnamefont {Zurek}},\ eds.,\
  \enquote {\bibinfo {title} {Quantum theory and measurement},}\ \ (\bibinfo
  {publisher} {Princeton University Press, Princeton, New Jersey
  1983})\BibitemShut {NoStop}%
\bibitem [{\citenamefont {{Kraus}}\ \emph {et~al.}(1983)\citenamefont
  {{Kraus}}, \citenamefont {{B{\"o}hm}}, \citenamefont {{Dollard}},\ and\
  \citenamefont {{Wootters}}}]{1983LNP...190.....K}%
  \BibitemOpen
  \bibinfo {editor} {\bibfnamefont {K.}~\bibnamefont {{Kraus}}}, \bibinfo
  {editor} {\bibfnamefont {A.}~\bibnamefont {{B{\"o}hm}}}, \bibinfo {editor}
  {\bibfnamefont {J.~D.}\ \bibnamefont {{Dollard}}}, \ and\ \bibinfo {editor}
  {\bibfnamefont {W.~H.}\ \bibnamefont {{Wootters}}},\ eds.,\ \href {\doibase
  10.1007/3-540-12732-1} {\emph {\bibinfo {title} {States, Effects, and
  Operations Fundamental Notions of Quantum Theory}}},\ \bibinfo {series}
  {Lecture Notes in Physics, Berlin Springer Verlag}, Vol.\ \bibinfo {volume}
  {190}\ (\bibinfo {year} {1983})\BibitemShut {NoStop}%
\bibitem [{\citenamefont {Busch}\ \emph {et~al.}(1995)\citenamefont {Busch},
  \citenamefont {Grabowski},\ and\ \citenamefont {Lahti}}]{busch-etal:1995}%
  \BibitemOpen
  \bibfield  {author} {\bibinfo {author} {\bibfnamefont {P.}~\bibnamefont
  {Busch}}, \bibinfo {author} {\bibfnamefont {M.}~\bibnamefont {Grabowski}}, \
  and\ \bibinfo {author} {\bibfnamefont {P.~J.}\ \bibnamefont {Lahti}},\ }in\
  \href {\doibase https://doi.org/10.1007/978-3-540-49239-9} {\emph {\bibinfo
  {booktitle} {{Operational Quantum Physics}}}},\ \bibinfo {series} {Lecture
  Notes in Physics Monographs}, Vol.~\bibinfo {volume} {31},\ \bibinfo {editor}
  {edited by\ \bibinfo {editor} {\bibnamefont {Springer}}}\ (\bibinfo
  {publisher} {Springer, Berlin, Heidelberg},\ \bibinfo {year} {1995})\
  Chap.~\bibinfo {chapter} {4}, pp.\ \bibinfo {pages} {95 -- 115}\BibitemShut
  {NoStop}%
\bibitem [{\citenamefont {{L{\"u}ders}}(2006)}]{2006AnP...518..663L}%
  \BibitemOpen
  \bibfield  {author} {\bibinfo {author} {\bibfnamefont {G.}~\bibnamefont
  {{L{\"u}ders}}},\ }\href {\doibase 10.1002/andp.200610207} {\bibfield
  {journal} {\bibinfo  {journal} {Annalen der Physik}\ }\textbf {\bibinfo
  {volume} {518}},\ \bibinfo {pages} {663} (\bibinfo {year} {2006})},\ \Eprint
  {http://arxiv.org/abs/quant-ph/0403007} {arXiv:quant-ph/0403007 [quant-ph]}
  \BibitemShut {NoStop}%
\bibitem [{Note1()}]{Note1}%
  \BibitemOpen
  \bibinfo {note} {There exist a lot of literatures which deals with the
  resource theory of operations, namely, Refs.\cite
  {PhysRevA.67.052301,PhysRevLett.122.190405,PhysRevA.95.062327,DATTA2018243}}\BibitemShut
  {NoStop}%
\bibitem [{\citenamefont {{Mauro D'Ariano}}\ \emph {et~al.}(2005)\citenamefont
  {{Mauro D'Ariano}}, \citenamefont {{Lo Presti}},\ and\ \citenamefont
  {{Perinotti}}}]{2005JPhA...38.5979M}%
  \BibitemOpen
  \bibfield  {author} {\bibinfo {author} {\bibfnamefont {G.}~\bibnamefont
  {{Mauro D'Ariano}}}, \bibinfo {author} {\bibfnamefont {P.}~\bibnamefont {{Lo
  Presti}}}, \ and\ \bibinfo {author} {\bibfnamefont {P.}~\bibnamefont
  {{Perinotti}}},\ }\href {\doibase 10.1088/0305-4470/38/26/010} {\bibfield
  {journal} {\bibinfo  {journal} {Journal of Physics A Mathematical General}\
  }\textbf {\bibinfo {volume} {38}},\ \bibinfo {pages} {5979} (\bibinfo {year}
  {2005})},\ \Eprint {http://arxiv.org/abs/quant-ph/0408115}
  {arXiv:quant-ph/0408115 [quant-ph]} \BibitemShut {NoStop}%
\bibitem [{\citenamefont {Johnston}(2016)}]{qetlab}%
  \BibitemOpen
  \bibfield  {author} {\bibinfo {author} {\bibfnamefont {N.}~\bibnamefont
  {Johnston}},\ }\href {\doibase 10.5281/zenodo.44637} {\enquote {\bibinfo
  {title} {{QETLAB}: A {MATLAB} toolbox for quantum entanglement, version
  0.9},}\ }\bibinfo {howpublished} {\url{http://qetlab.com}} (\bibinfo {year}
  {2016})\BibitemShut {NoStop}%
\bibitem [{\citenamefont {{Heinosaari}}\ \emph {et~al.}(2019)\citenamefont
  {{Heinosaari}}, \citenamefont {{Anastasia Jivulescu}},\ and\ \citenamefont
  {{Nechita}}}]{2019arXiv190204751H}%
  \BibitemOpen
  \bibfield  {author} {\bibinfo {author} {\bibfnamefont {T.}~\bibnamefont
  {{Heinosaari}}}, \bibinfo {author} {\bibfnamefont {M.}~\bibnamefont
  {{Anastasia Jivulescu}}}, \ and\ \bibinfo {author} {\bibfnamefont
  {I.}~\bibnamefont {{Nechita}}},\ }\href@noop {} {\bibfield  {journal}
  {\bibinfo  {journal} {arXiv e-prints}\ ,\ \bibinfo {eid} {arXiv:1902.04751}}
  (\bibinfo {year} {2019})},\ \Eprint {http://arxiv.org/abs/1902.04751}
  {arXiv:1902.04751 [quant-ph]} \BibitemShut {NoStop}%
\bibitem [{\citenamefont {Ahnert}\ and\ \citenamefont
  {Payne}(2005)}]{PhysRevA.71.012330}%
  \BibitemOpen
  \bibfield  {author} {\bibinfo {author} {\bibfnamefont {S.~E.}\ \bibnamefont
  {Ahnert}}\ and\ \bibinfo {author} {\bibfnamefont {M.~C.}\ \bibnamefont
  {Payne}},\ }\href {\doibase 10.1103/PhysRevA.71.012330} {\bibfield  {journal}
  {\bibinfo  {journal} {Phys. Rev. A}\ }\textbf {\bibinfo {volume} {71}},\
  \bibinfo {pages} {012330} (\bibinfo {year} {2005})}\BibitemShut {NoStop}%
\bibitem [{\citenamefont {Paris}(2012)}]{Paris2012}%
  \BibitemOpen
  \bibfield  {author} {\bibinfo {author} {\bibfnamefont {M.~G.~A.}\
  \bibnamefont {Paris}},\ }\href {\doibase 10.1140/epjst/e2012-01535-1}
  {\bibfield  {journal} {\bibinfo  {journal} {The European Physical Journal
  Special Topics}\ }\textbf {\bibinfo {volume} {203}},\ \bibinfo {pages} {61}
  (\bibinfo {year} {2012})}\BibitemShut {NoStop}%
\bibitem [{\citenamefont {Rana}\ \emph {et~al.}(2016)\citenamefont {Rana},
  \citenamefont {Parashar},\ and\ \citenamefont
  {Lewenstein}}]{PhysRevA.93.012110}%
  \BibitemOpen
  \bibfield  {author} {\bibinfo {author} {\bibfnamefont {S.}~\bibnamefont
  {Rana}}, \bibinfo {author} {\bibfnamefont {P.}~\bibnamefont {Parashar}}, \
  and\ \bibinfo {author} {\bibfnamefont {M.}~\bibnamefont {Lewenstein}},\
  }\href {\doibase 10.1103/PhysRevA.93.012110} {\bibfield  {journal} {\bibinfo
  {journal} {Phys. Rev. A}\ }\textbf {\bibinfo {volume} {93}},\ \bibinfo
  {pages} {012110} (\bibinfo {year} {2016})}\BibitemShut {NoStop}%
\bibitem [{Note2()}]{Note2}%
  \BibitemOpen
  \bibinfo {note} {It quantifies the \# of Bell states one can extract from $N$
  number of mixed entangled states. For further study, we refer readers to
  \cite {PhysRevA.53.2046}.}\BibitemShut {Stop}%
\bibitem [{Note3()}]{Note3}%
  \BibitemOpen
  \bibinfo {note} {The work deficit is the information, or work, that cannot be
  extracted from a bipartite quantum state when the parties are in distant
  locations, as compared to the case when the same are together. For more
  information on work deficit, we direct the enthusiastic reader to Refs. \cite
  {PhysRevLett.89.180402,PhysRevLett.90.100402,PhysRevA.71.062307}}\BibitemShut
  {NoStop}%
\bibitem [{\citenamefont {Streltsov}\ \emph {et~al.}(2011)\citenamefont
  {Streltsov}, \citenamefont {Kampermann},\ and\ \citenamefont
  {Bru\ss{}}}]{Link-Disc}%
  \BibitemOpen
  \bibfield  {author} {\bibinfo {author} {\bibfnamefont {A.}~\bibnamefont
  {Streltsov}}, \bibinfo {author} {\bibfnamefont {H.}~\bibnamefont
  {Kampermann}}, \ and\ \bibinfo {author} {\bibfnamefont {D.}~\bibnamefont
  {Bru\ss{}}},\ }\href {\doibase 10.1103/PhysRevLett.106.160401} {\bibfield
  {journal} {\bibinfo  {journal} {Phys. Rev. Lett.}\ }\textbf {\bibinfo
  {volume} {106}},\ \bibinfo {pages} {160401} (\bibinfo {year}
  {2011})}\BibitemShut {NoStop}%
\bibitem [{\citenamefont {Piani}\ \emph {et~al.}(2011)\citenamefont {Piani},
  \citenamefont {Gharibian}, \citenamefont {Adesso}, \citenamefont
  {Calsamiglia}, \citenamefont {Horodecki},\ and\ \citenamefont
  {Winter}}]{PhysRevLett.106.220403}%
  \BibitemOpen
  \bibfield  {author} {\bibinfo {author} {\bibfnamefont {M.}~\bibnamefont
  {Piani}}, \bibinfo {author} {\bibfnamefont {S.}~\bibnamefont {Gharibian}},
  \bibinfo {author} {\bibfnamefont {G.}~\bibnamefont {Adesso}}, \bibinfo
  {author} {\bibfnamefont {J.}~\bibnamefont {Calsamiglia}}, \bibinfo {author}
  {\bibfnamefont {P.}~\bibnamefont {Horodecki}}, \ and\ \bibinfo {author}
  {\bibfnamefont {A.}~\bibnamefont {Winter}},\ }\href {\doibase
  10.1103/PhysRevLett.106.220403} {\bibfield  {journal} {\bibinfo  {journal}
  {Phys. Rev. Lett.}\ }\textbf {\bibinfo {volume} {106}},\ \bibinfo {pages}
  {220403} (\bibinfo {year} {2011})}\BibitemShut {NoStop}%
\bibitem [{\citenamefont {Nielsen}\ \emph {et~al.}(2003)\citenamefont
  {Nielsen}, \citenamefont {Dawson}, \citenamefont {Dodd}, \citenamefont
  {Gilchrist}, \citenamefont {Mortimer}, \citenamefont {Osborne}, \citenamefont
  {Bremner}, \citenamefont {Harrow},\ and\ \citenamefont
  {Hines}}]{PhysRevA.67.052301}%
  \BibitemOpen
  \bibfield  {author} {\bibinfo {author} {\bibfnamefont {M.~A.}\ \bibnamefont
  {Nielsen}}, \bibinfo {author} {\bibfnamefont {C.~M.}\ \bibnamefont {Dawson}},
  \bibinfo {author} {\bibfnamefont {J.~L.}\ \bibnamefont {Dodd}}, \bibinfo
  {author} {\bibfnamefont {A.}~\bibnamefont {Gilchrist}}, \bibinfo {author}
  {\bibfnamefont {D.}~\bibnamefont {Mortimer}}, \bibinfo {author}
  {\bibfnamefont {T.~J.}\ \bibnamefont {Osborne}}, \bibinfo {author}
  {\bibfnamefont {M.~J.}\ \bibnamefont {Bremner}}, \bibinfo {author}
  {\bibfnamefont {A.~W.}\ \bibnamefont {Harrow}}, \ and\ \bibinfo {author}
  {\bibfnamefont {A.}~\bibnamefont {Hines}},\ }\href {\doibase
  10.1103/PhysRevA.67.052301} {\bibfield  {journal} {\bibinfo  {journal} {Phys.
  Rev. A}\ }\textbf {\bibinfo {volume} {67}},\ \bibinfo {pages} {052301}
  (\bibinfo {year} {2003})}\BibitemShut {NoStop}%
\bibitem [{\citenamefont {Theurer}\ \emph {et~al.}(2019)\citenamefont
  {Theurer}, \citenamefont {Egloff}, \citenamefont {Zhang},\ and\ \citenamefont
  {Plenio}}]{PhysRevLett.122.190405}%
  \BibitemOpen
  \bibfield  {author} {\bibinfo {author} {\bibfnamefont {T.}~\bibnamefont
  {Theurer}}, \bibinfo {author} {\bibfnamefont {D.}~\bibnamefont {Egloff}},
  \bibinfo {author} {\bibfnamefont {L.}~\bibnamefont {Zhang}}, \ and\ \bibinfo
  {author} {\bibfnamefont {M.~B.}\ \bibnamefont {Plenio}},\ }\href {\doibase
  10.1103/PhysRevLett.122.190405} {\bibfield  {journal} {\bibinfo  {journal}
  {Phys. Rev. Lett.}\ }\textbf {\bibinfo {volume} {122}},\ \bibinfo {pages}
  {190405} (\bibinfo {year} {2019})}\BibitemShut {NoStop}%
\bibitem [{\citenamefont {Ben~Dana}\ \emph {et~al.}(2017)\citenamefont
  {Ben~Dana}, \citenamefont {Garc\'{\i}a~D\'{\i}az}, \citenamefont {Mejatty},\
  and\ \citenamefont {Winter}}]{PhysRevA.95.062327}%
  \BibitemOpen
  \bibfield  {author} {\bibinfo {author} {\bibfnamefont {K.}~\bibnamefont
  {Ben~Dana}}, \bibinfo {author} {\bibfnamefont {M.}~\bibnamefont
  {Garc\'{\i}a~D\'{\i}az}}, \bibinfo {author} {\bibfnamefont {M.}~\bibnamefont
  {Mejatty}}, \ and\ \bibinfo {author} {\bibfnamefont {A.}~\bibnamefont
  {Winter}},\ }\href {\doibase 10.1103/PhysRevA.95.062327} {\bibfield
  {journal} {\bibinfo  {journal} {Phys. Rev. A}\ }\textbf {\bibinfo {volume}
  {95}},\ \bibinfo {pages} {062327} (\bibinfo {year} {2017})}\BibitemShut
  {NoStop}%
\bibitem [{\citenamefont {Datta}\ \emph {et~al.}(2018)\citenamefont {Datta},
  \citenamefont {Sazim}, \citenamefont {Pati},\ and\ \citenamefont
  {Agrawal}}]{DATTA2018243}%
  \BibitemOpen
  \bibfield  {author} {\bibinfo {author} {\bibfnamefont {C.}~\bibnamefont
  {Datta}}, \bibinfo {author} {\bibfnamefont {S.}~\bibnamefont {Sazim}},
  \bibinfo {author} {\bibfnamefont {A.~K.}\ \bibnamefont {Pati}}, \ and\
  \bibinfo {author} {\bibfnamefont {P.}~\bibnamefont {Agrawal}},\ }\href
  {\doibase https://doi.org/10.1016/j.aop.2018.08.014} {\bibfield  {journal}
  {\bibinfo  {journal} {Annals of Physics}\ }\textbf {\bibinfo {volume}
  {397}},\ \bibinfo {pages} {243 } (\bibinfo {year} {2018})}\BibitemShut
  {NoStop}%
\bibitem [{\citenamefont {Bennett}\ \emph {et~al.}(1996)\citenamefont
  {Bennett}, \citenamefont {Bernstein}, \citenamefont {Popescu},\ and\
  \citenamefont {Schumacher}}]{PhysRevA.53.2046}%
  \BibitemOpen
  \bibfield  {author} {\bibinfo {author} {\bibfnamefont {C.~H.}\ \bibnamefont
  {Bennett}}, \bibinfo {author} {\bibfnamefont {H.~J.}\ \bibnamefont
  {Bernstein}}, \bibinfo {author} {\bibfnamefont {S.}~\bibnamefont {Popescu}},
  \ and\ \bibinfo {author} {\bibfnamefont {B.}~\bibnamefont {Schumacher}},\
  }\href {\doibase 10.1103/PhysRevA.53.2046} {\bibfield  {journal} {\bibinfo
  {journal} {Phys. Rev. A}\ }\textbf {\bibinfo {volume} {53}},\ \bibinfo
  {pages} {2046} (\bibinfo {year} {1996})}\BibitemShut {NoStop}%
\bibitem [{\citenamefont {Oppenheim}\ \emph {et~al.}(2002)\citenamefont
  {Oppenheim}, \citenamefont {Horodecki}, \citenamefont {Horodecki},\ and\
  \citenamefont {Horodecki}}]{PhysRevLett.89.180402}%
  \BibitemOpen
  \bibfield  {author} {\bibinfo {author} {\bibfnamefont {J.}~\bibnamefont
  {Oppenheim}}, \bibinfo {author} {\bibfnamefont {M.}~\bibnamefont
  {Horodecki}}, \bibinfo {author} {\bibfnamefont {P.}~\bibnamefont
  {Horodecki}}, \ and\ \bibinfo {author} {\bibfnamefont {R.}~\bibnamefont
  {Horodecki}},\ }\href {\doibase 10.1103/PhysRevLett.89.180402} {\bibfield
  {journal} {\bibinfo  {journal} {Phys. Rev. Lett.}\ }\textbf {\bibinfo
  {volume} {89}},\ \bibinfo {pages} {180402} (\bibinfo {year}
  {2002})}\BibitemShut {NoStop}%
\bibitem [{\citenamefont {Horodecki}\ \emph {et~al.}(2003)\citenamefont
  {Horodecki}, \citenamefont {Horodecki}, \citenamefont {Horodecki},
  \citenamefont {Horodecki}, \citenamefont {Oppenheim}, \citenamefont
  {Sen(De)},\ and\ \citenamefont {Sen}}]{PhysRevLett.90.100402}%
  \BibitemOpen
  \bibfield  {author} {\bibinfo {author} {\bibfnamefont {M.}~\bibnamefont
  {Horodecki}}, \bibinfo {author} {\bibfnamefont {K.}~\bibnamefont
  {Horodecki}}, \bibinfo {author} {\bibfnamefont {P.}~\bibnamefont
  {Horodecki}}, \bibinfo {author} {\bibfnamefont {R.}~\bibnamefont
  {Horodecki}}, \bibinfo {author} {\bibfnamefont {J.}~\bibnamefont
  {Oppenheim}}, \bibinfo {author} {\bibfnamefont {A.}~\bibnamefont {Sen(De)}},
  \ and\ \bibinfo {author} {\bibfnamefont {U.}~\bibnamefont {Sen}},\ }\href
  {\doibase 10.1103/PhysRevLett.90.100402} {\bibfield  {journal} {\bibinfo
  {journal} {Phys. Rev. Lett.}\ }\textbf {\bibinfo {volume} {90}},\ \bibinfo
  {pages} {100402} (\bibinfo {year} {2003})}\BibitemShut {NoStop}%
\bibitem [{\citenamefont {Horodecki}\ \emph {et~al.}(2005)\citenamefont
  {Horodecki}, \citenamefont {Horodecki}, \citenamefont {Horodecki},
  \citenamefont {Oppenheim}, \citenamefont {Sen(De)}, \citenamefont {Sen},\
  and\ \citenamefont {Synak-Radtke}}]{PhysRevA.71.062307}%
  \BibitemOpen
  \bibfield  {author} {\bibinfo {author} {\bibfnamefont {M.}~\bibnamefont
  {Horodecki}}, \bibinfo {author} {\bibfnamefont {P.}~\bibnamefont
  {Horodecki}}, \bibinfo {author} {\bibfnamefont {R.}~\bibnamefont
  {Horodecki}}, \bibinfo {author} {\bibfnamefont {J.}~\bibnamefont
  {Oppenheim}}, \bibinfo {author} {\bibfnamefont {A.}~\bibnamefont {Sen(De)}},
  \bibinfo {author} {\bibfnamefont {U.}~\bibnamefont {Sen}}, \ and\ \bibinfo
  {author} {\bibfnamefont {B.}~\bibnamefont {Synak-Radtke}},\ }\href {\doibase
  10.1103/PhysRevA.71.062307} {\bibfield  {journal} {\bibinfo  {journal} {Phys.
  Rev. A}\ }\textbf {\bibinfo {volume} {71}},\ \bibinfo {pages} {062307}
  (\bibinfo {year} {2005})}\BibitemShut {NoStop}%
\bibitem [{\citenamefont {Zhengjun}\ \emph {et~al.}(2014)\citenamefont
	{Zhengjun}, \citenamefont {Yongming},\ and\ \citenamefont
	{Heng}}]{Xi2015}%
\BibitemOpen
\bibfield  {author} {\bibinfo {author} {\bibfnamefont {Xi.}~\bibnamefont
		{Zhengjun}}, \bibinfo {author} {\bibfnamefont {Li.}~\bibnamefont {Yongming}}, \
	and\ \bibinfo {author} {\bibfnamefont {Fan.}\ \bibnamefont {Heng}},\
}\href {\doibase 10.1038/srep10922} {\bibfield  {journal}
	{\bibinfo  {journal} {Scientific Reports}\ }\textbf {\bibinfo {volume} {5}},\
	\bibinfo {pages} {10922} (\bibinfo {year} {2015})}\BibitemShut {NoStop}%
\end{thebibliography}
\end{document}